\documentclass{jfm}
\usepackage{graphicx}
\usepackage{epstopdf, epsfig}
\usepackage{amsmath}
\usepackage{booktabs}
\usepackage{float}
\usepackage{color}
\usepackage{xcolor}

\shorttitle{Viscometric flow of dense granular materials}
\shortauthor{I. Srivastava, L. E. Silbert, G. S. Grest and J. B. Lechman}

\title{Viscometric flow of dense granular materials under controlled pressure and shear stress}

\author{Ishan Srivastava\aff{1,2}
  \corresp{\email{isriva@lbl.gov}},
  Leonardo E. Silbert\aff{3},
  Gary S. Grest\aff{1},
 \and Jeremy B. Lechman\aff{1}
 \corresp{\email{jblechm@sandia.gov}}}

\affiliation{\aff{1}Sandia National Laboratories,
Albuquerque, NM 87185, USA
\aff{2}Center for Computational Sciences and Engineering, 
Lawrence Berkeley National Laboratory, Berkeley, CA 94720, USA
\aff{3}School of Math, Science, and Engineering, Central New Mexico Community College,
Albuquerque, NM 87106, USA}

\begin{document}

\maketitle

\begin{abstract}
This study examines the flow of dense granular materials under external shear stress and pressure using discrete element method simulations. In this  method, the material is allowed to strain along all periodic directions and adapt its solid volume fraction in response to an imbalance between the internal state of stress and the external applied stress. By systematically varying the external shear stress and pressure, the steady rheological response is simulated for: (1) rate-independent quasi-static flow, and (2) rate-dependent inertial flow. The simulated flow is viscometric with non-negligible first and second normal stress differences. While both normal stress differences are negative in inertial flows, the first normal stress difference switches from negative to slightly positive, and second normal stress difference tends to zero in quasi-static flows. The first normal stress difference emerges from a lack of co-axiality between a second-rank contact fabric tensor and strain rate tensor in the flow plane, while the second normal stress difference is linked to an excess of contacts in the shear plane compared to the vorticity direction. A general rheological model of second order (in terms of strain rate tensor) is proposed to describe the two types of flow, and the model is calibrated for various values of interparticle friction from simulations on nearly mono-disperse spheres. The model incorporates normal stress differences in both regimes of flow and provides a complete viscometric description of steady dense granular flows.
\end{abstract}


\section{Introduction}\label{intro}
Granular flows exhibit several intriguing phenomena that distinguish them from Newtonian fluids, such as the presence of pressure-dependent arrest and flow onset (yield) criteria leading to rate-independent and rate-dependent flows, and a dilute gas-like flow dominated by inelastic particle collisions. A convenient classification defines three distinct types of granular flows~\citep{forterre2008}: (1) quasi-static flows, (2) dense inertial flows, and (3) gas-like collisional flows. The behavior of the three types of granular flows is quite diverse and several constitutive models have been proposed for their description; however, a general constitutive model applicable across all flow types remains elusive. In this work we focus our attention on the rate-independent and rate-dependent flows, where the particle contact lifetimes are relatively long, inertia is important and the material is predominantly dense. 

The rate-independent flow regime has been described by various constitutive models, largely inspired by the principles of solid mechanics and plasticity. Beginning with the incipient failure hypothesis of the Coulomb yield criterion~\citep{sokolovskii2016statics}, further observations of critical state deformation in soils led to the development of several rigid-plastic and elasto-plastic models based on critical state theory and associated plasticity~\citep{schofield1968critical}. Recent developments have attempted to include material anisotropy in granular plasticity by introducing state-dependence of material stress, often characterized via material texture or fabric~\citep{li2012,gao2014,sun2011}. The rate-independent granular plasticity has also been characterized by double shearing models~\citep{spencer1964} that relax the assumption of homogeneous deformations to explain shear banding along slip planes in granular materials. Further advances on these models have introduced the concepts of dilatancy~\citep{mehrabadi1978}, work hardening~\citep{anand2000}, and more recently fabric anisotropy~\citep{nemat-nasser2000,zhu2006}. The reader is referred to a recent review of various constitutive models of rate-independent regime in granular flows~\citep{radjai2017}. 

Rate-dependent granular flows were first observed to exhibit quadratic scaling of shear and normal stress with strain rate at a constant volume~\citep{bagnold1954}, which was later verified in several experiments and simulations~\citep{dacruz2005,lois2005,pouliquen1999,silbert2001}. Recently, a Bingham-type $\mu(I)$ rheological model for granular materials at moderate shear rates has been proposed~\citep{jop2006}, which attempts to connect the rate-independent and rate-dependent granular flow regimes by introducing pressure as a control variable instead of volume, although the two can be interchanged based on the one-to-one relationship between $\mu$ and $I$ at moderate shear rates. Here $\mu$ is the dimensionless shear stress ratio, and $I$ is the dimensionless inertial number (described later in the text). A detailed discussion of such viscoplastic models can be found in a recent review~\citep{goddard2014}.

Although these rheological models have successfully predicted granular flow profiles in a remarkable number of geometries~\citep{midi2004}, several rheological effects remain unexplained, such as surface curvature in free-surface flows~\citep{couturier2011,mcelwaine2012}, negative rod climbing effects~\citep{boyer2011}, anomalous stress profile in Couette flows~\citep{mehandia2012}, and the observation of shear-free sheets in split-bottom Couette flows~\citep{depken2007}. Many of these effects arise from a lack of co-axiality between principal directions of stress and strain rate tensors in viscometric flows~\citep{alam2003,alam2005,rycroft2009,weinhart2013,saha2016,seto2018,bhateja2018}, which is the operating assumption in several constitutive models. As such, there is a need for higher-order constitutive models that incorporate these effects to provide better predictions of granular rheology. Furthermore, microstructural origins of these rheological effects, especially in the dense and quasi-static flow regimes, remain unclear.  

In this paper, we describe fully stress-controlled discrete element method (DEM) simulations in both rate-independent and rate-dependent regimes. This simulation method enables the evolution of all strain degrees of freedom of a fully-periodic representative volume element of granular material in response to external applied shear stress and pressure. The novelty of this simulation method is four-fold: (1) it naturally captures the pressure-dependent flow-onset (yield) and flow-arrest phenomena~\citep{srivastava2019}, (2) by prescribing shear stress rather than shear rate, this method can seamlessly traverse across rate-dependent and rate-independent regimes of granular flow, (3) by prescribing pressure rather than solid volume fraction, shear-induced dilation of granular materials is fully captured, and (4) the fully periodic nature of the simulations is devoid of any boundary effects, and thus represents the true bulk response of granular materials to applied stresses. The stress-controlled method is used to simulate shear flows of nearly mono-disperse spheres. A second-order rheological model that does not assume co-axiality of stress and strain rate tensors is proposed, and the model is calibrated for various values of interparticle friction from the simulation data. For viscometric flows, the second-order rheological effects result in non-negligible first and normal stress differences. We provide microstructural insights into the origin of normal stress differences by analyzing a second-rank contact fabric tensor.

The paper is organized as follows. Section 2 on Model and Methods describes the second-order rheological model, introduces the stress-controlled DEM simulation method, and provides details on the sphere-sphere contact mechanics model and general simulation parameters. Section 3 provides evidence that the steady flow generated in these simulations by applying shear stress and pressure is viscometric in nature. Section 4 describes the calibration of the rheological model based on the viscometric flow data from DEM simulations. Section 5 describes the normal stress differences measured in these simulations and their microstructural origins.

\section{Model and Methods}\label{model}
\subsection{Rheological Model}
In anticipation of the results presented below, we introduce a purely-dissipative rheological framework proposed by \citet{goddard1984}, which was utilized to formulate constitutive laws for rate-independent and rate-dependent flow in granular materials~\citep{goddard1986,goddard2014}. In this framework, stress emerges from dissipation through macroscopic bulk deformation, which dominates over grain-scale inertial relaxation. This is similar to the dense flow of granular materials at low inertial numbers, which is of interest here. Furthermore, elastic effects are ignored, and non-hydrostatic stress components emerge entirely from granular flow, and are zero when there is no flow. In this framework, the Cauchy stress tensor $\boldsymbol{\sigma}$ is given by:
\begin{equation}
    \boldsymbol{\sigma} = \boldsymbol{\mathcal{\eta}}\{\mathcal{H}\}:\boldsymbol{D},
    \label{eq1a}
\end{equation}
where $\boldsymbol{D}$ is the symmetric strain rate tensor, and $\boldsymbol{\mathcal{\eta}}\{\mathcal{H}\}$ is a positive-definite fourth-rank tensor adhering to the constraints of a purely dissipative material, i.e., $\boldsymbol{D}\!:\!\boldsymbol{\mathcal{\eta}}\{\mathcal{H}\}\!:\!\boldsymbol{D}\!>\!0$, and is dependent upon the history $\mathcal{H}$ of flow, which can be conveniently represented through a deformation gradient $\boldsymbol{F}$ relative to a reference state. For the specific case of $|\boldsymbol{D}|\to0$ corresponding to rate-independent plastic deformation (here, $|\boldsymbol{D}|\!=\!\sqrt{\frac{1}{2}\boldsymbol{D}\!:\!\boldsymbol{D}}$), the stress is given as:
\begin{equation}
    \boldsymbol{\sigma} = \frac{\boldsymbol{\mu}_{0}\{\mathcal{H}\}:\boldsymbol{D}}{|\boldsymbol{D}|},
    \label{eq1b}
\end{equation}
where $\boldsymbol{\mu}_{0}\{\mathcal{H}\}$ is a fourth-rank \textit{yield modulus}. Therefore, the total stress can be partitioned into its rate-independent and rate-dependent components as:
\begin{equation}
    \boldsymbol{\sigma} = \frac{\boldsymbol{\mu}_{0}\{\mathcal{H}\}:\boldsymbol{D}}{|\boldsymbol{D}|} + \boldsymbol{\mathcal{\eta}}_{0}\{\mathcal{H}\}:\boldsymbol{D},
    \label{eq1c}
\end{equation}
where $\boldsymbol{\mathcal{\eta}}_{0}\{\mathcal{H}\}$ is a fourth-rank \textit{viscosity} tensor. 

We adapt this rheological framework to model granular rheology through the following assumptions that will be demonstrated to hold true in the present simulations: (1) the flow is homogeneous with a constant stretch history~\citep{noll1962}, and (2) the flow is planar and isochoric, i.e., $\boldsymbol{D}$ is characterized by two dominant eigenvalues and $\mathrm{tr}(\boldsymbol{D})=0$. This dependence is introduced in a frame-indifferent manner to produce a second-order rheological model that well-describes non-isotropic flow effects observed in our simulations. With these simplifications, $\boldsymbol{\sigma}$ can be represented as:
\begin{eqnarray}
    \boldsymbol{\sigma} &=& p\boldsymbol{I}+\eta_{1}\boldsymbol{D}+\eta_{2}\left[\boldsymbol{D}^{2}-\frac{\mathrm{tr}\left(\boldsymbol{D}^{2}\right)}{3}\boldsymbol{I}\right]+\eta_{3}\left[\dot{\boldsymbol{D}} - \boldsymbol{W}\boldsymbol{D}+\boldsymbol{D}\boldsymbol{W}\right]\nonumber \\
    &&+\kappa_{1}\frac{\boldsymbol{D}}{|\boldsymbol{D}|}+\kappa_{2}\left[\frac{\boldsymbol{D}^{2}}{|\boldsymbol{D}|^{2}}-\frac{\mathrm{tr}\left(\boldsymbol{D}^{2}\right)}{3|\boldsymbol{D}|^{2}}\boldsymbol{I}\right], 
    \label{eq1}
\end{eqnarray}
where 
\begin{equation}
    \dot{\boldsymbol{D}}=\frac{\partial \boldsymbol{D}}{\partial t}+\boldsymbol{v}\cdot\nabla\boldsymbol{D}
    \label{eq1d}
\end{equation}
is the material derivative of $\boldsymbol{D}$, $\boldsymbol{v}$ is the material velocity, and $\mathring{\boldsymbol{D}}\!=\!\dot{\boldsymbol{D}}\!-\!\boldsymbol{W}\boldsymbol{D}\!+\!\boldsymbol{D}\boldsymbol{W}$ represents the frame-indifferent co-rotational derivative of $\boldsymbol{D}$~\citep{bird1987dynamics}. The isotropic pressure is defined as $p=\frac{1}{3}\mathrm{tr}\left(\boldsymbol{\sigma}\right)$, and $\boldsymbol{I}$ is the unit tensor. The deviatoric stress, $\boldsymbol{\sigma} - p\boldsymbol{I}$, depends on $\boldsymbol{D}=1/2(\nabla \boldsymbol{v} + \nabla \boldsymbol{v}^{T})$ and a vorticity tensor $\boldsymbol{W}=1/2(\nabla \boldsymbol{v} - \nabla \boldsymbol{v}^{T})$. Here, $\dot{\gamma}=|\boldsymbol{D}|$ is the magnitude of the strain-rate tensor. The second, third and fourth terms in (\ref{eq1}) represent rate-dependent contributions to the total stress that are characterized by the flow functions $\eta_{1}(\dot{\gamma},p)$ and $\eta_{2}(\dot{\gamma},p)$ and $\eta_{3}(\dot{\gamma},p)$, and are similar in form to a second-order description of non-Newtonian fluids using Rivlin-Erickson tensors~\citep{rivlin1955stress}. The fifth and sixth terms in (\ref{eq1}) represent rate-independent contributions to the total stress that are characterized by plastic yield-like functions $\kappa_{1}(p)$ and $\kappa_{2}(p)$, which generally depend on the flow history. The pressure dependence of the flow functions is similar in spirit to the implicit constitutive theory of~\citet{rajagopal2006}. In this work we focus on simple shear flows, but in general, the coefficients $\eta_{1}$, $\eta_{2}$ and $\eta_{3}$ depend on $\mathrm{tr}(\boldsymbol{D}^2)$ and $\mathrm{tr}(\mathring{\boldsymbol{D}}^2)$, which is important when modeling non-viscometric flows \citep{giusteri2018}. Similarly, the coefficients $\kappa_{1}$ and $\kappa_{2}$ depend on $\mathrm{tr}(\boldsymbol{D}^2)/|\boldsymbol{D}|^{2}$, which can be calibrated from anisotropic models of granular yield criterion. Such anisotropy was demonstrated in simulations~\citep{thornton2010,li2012}, resulting in deviations from the Drucker-Prager like isotropic yield criterion that is implicit in the $\mu(I)$ rheology~\citep{jop2006}. Furthermore, the rheological model can be extended to multi-axial flows that are observed in practice~\citep{cortet2009relevance} by introducing additional dependence of flow functions on $\mathrm{tr}(\boldsymbol{D}^3)$ and $\mathrm{tr}(\mathring{\boldsymbol{D}}^3)$~\citep{wang1965representation,larson1985flows}.

In this paper, we will consider steady homogeneous planar shear flow of granular materials resulting from a constant applied external shear stress and pressure, in which the memory of the flow has decayed and the deformation history is unimportant. In such steady homogeneous flows $\dot{\boldsymbol{D}}=0$, indicating that the eigenvectors of $\boldsymbol{D}$ are uniform in space and time, and local material rotation arises entirely from flow vorticity~\citep{schunk1990,giusteri2018}. Consider a uniform velocity gradient $\nabla \boldsymbol{v}$ with the following viscometric form:
\begin{equation}
    \nabla \boldsymbol{v} = \left[
\begin{array}{ccc}
  0 & 2\dot{\gamma} & 0  \\
  0 & 0 & 0 \\
  0 & 0 & 0 \\
\end{array}  \right],
\label{eq2}
\end{equation}
for flow along $x$ direction, velocity gradient along $y$ direction, and vorticity along $z$ direction, and where $\mathrm{tr}(\boldsymbol{D}^3)=0$. In such viscometric flows, $\eta_{1}$, $\eta_{2}$, and $\eta_{3}$ represent the standard viscometric flow functions for non-Newtonian fluids~\citep{coleman2012viscometric} corresponding to shear stress, second normal stress difference and first normal difference respectively. Similarly, $\kappa_{1}$ and $\kappa_{2}$ represent the analogous rate-independent flow functions. Correspondingly, for such viscometric flows, the stress tensor takes the following general form:
\begin{equation}
    \boldsymbol{\sigma} = \left[
\begin{array}{ccc}
  \sigma_{xx} & \sigma_{xy} & 0  \\
  \sigma_{xy} & \sigma_{yy} & 0 \\
  0 & 0 & \sigma_{zz} \\
\end{array}  \right],
\label{eq2a}
\end{equation}
where $\sigma_{xx}\neq\sigma_{yy}\neq\sigma_{zz}$. Previous simulations on sheared granular flows have proposed a similar form for the stress tensor, such as in granular flows down an incline~\citep{silbert2001,weinhart2013}, free surface flows~\citep{mcelwaine2012}, and in shear-free sheets  model~\citep{depken2006} that proposed $\sigma_{xx}=\sigma_{yy}\neq\sigma_{zz}$ for quasi-static granular flows in a split-bottom Couette cell~\citep{depken2007}.

The rheological model reduces to the well-known $\mu(I)$ relationship~\citep{jop2006} for sheared granular flows when the second-order coefficients $\eta_{2,3}=0$ and $\kappa_{2}=0$. In this case, the stress tensor is assumed to be co-axial with the strain rate tensor, and the two are related to each other by a scalar relationship:
\begin{equation}
    \boldsymbol{\sigma}=p\boldsymbol{I}+\mu(I)p\frac{\boldsymbol{D}}{|\boldsymbol{D}|},
    \label{eq3}
\end{equation}
where the stress ratio $\mu=|\boldsymbol{\sigma}-p\boldsymbol{I}|/p$ and the inertial number $I=|\boldsymbol{D}|a/(p/\rho)^{0.5}$, for an average particle diameter $a$ and material density $\rho$. The $\mu(I)$ function is related to the flow coefficients of the rheological model in (\ref{eq1}) through:
\begin{equation}
    \mu(I) = \frac{1}{p}\left(\eta_{1}|\boldsymbol{D}|+\kappa_{1}\right)
    \label{eq4}
\end{equation}

\subsection{Constant Stress Simulations}
Steady sheared flows can be simulated by applying a constant strain rate or a constant stress on the granular material. Previous simulations on granular flows have imposed a constant strain rate either through a solid wall-driven flow~\citep{dacruz2005,koval2009,kamrin2014,salerno2018} or by shearing the periodic simulation domain~\citep{campbell2002,campbell2005,otsuki2011,sun2011,srivastava2019}. Wall-driven granular flows often result in flow localization near the walls~\citep{shojaaee2012}, which requires careful calibration of wall properties to extract the bulk rheological properties~\citep{shojaaee2012a,schuhmacher2017wall}. While a constant strain rate at the periodic boundaries can produce a viscometric flow field without walls~\citep{campbell2002,campbell2005,peyneau2008,sun2011}, it often results in large shear stress fluctuations~\citep{peyneau2008}, especially in the quasi-static flow regime, which makes it challenging to calibrate the rate-independent part of granular rheology. Additionally, it was recently demonstrated that near the critical yield stress, granular flows are highly intermittent with a stochastic flow-arrest transition behavior~\citep{srivastava2019}. As such, a constant stress boundary condition is able to provide an accurate prediction of the rheology near the yield stress~\citep{srivastava2019}. In this work, we simulate granular flows by applying a constant shear stress at the periodic boundaries, in which material is allowed to flow or not depending on the magnitude of applied stress. We will demonstrate that this boundary condition results in a well-defined viscometric flow.

Granular flows can also be simulated either at constant volume (isochoric) ~\citep{campbell2002,sun2011,otsuki2011} or by imposing a constant normal stress~\citep{campbell2005,sun2011,favierdecoulomb2017,srivastava2019}. Granular materials dilate upon shearing, resulting in significant differences in the rheology between the two conditions~\citep{campbell2005}. When the applied normal stress is constant, the material can dilate or compact upon shearing (depending on the initial condition) towards a `critical state' solid volume fraction in the quasi-static regime~\citep{schofield1968critical,srivastava2019}. Furthermore, granular materials exhibit shear-induced dilation in the inertial regime. Isochoric granular flows are not commonly observed in practice, and various experiments often naturally correspond to a constant normal stress condition, such as in free surface flows~\citep{mcelwaine2012,jop2006} or flows in Couette cells ~\citep{lu2007,dijksman2011}. Furthermore, a constant volume condition precludes the possibility of simulating granular flows near the yield stress in the quasi-static regime. If the solid volume fraction is set lower than the critical solid volume fraction at the onset of flow, then a $\mu(I)$ frictional rheology can not be extracted as the shear stress will go to zero (rather than its yield threshold value) as the strain rate goes to zero. Similarly, if the volume fraction is set greater than its critical value, the flow is prohibited for any applied stress in the limit of rigid grains. In this work, we simulate granular flows at a constant applied pressure where the material is free to adapt its volume. A constant pressure condition is different from the case where all the normal stress components are specified equal to each other, as simulated previously in \citet{peyneau2008}. This allows an efficient estimation of normal stress differences that will be described later in the text.

To simulate the evolution of a granular system under constant external stress and pressure, we utilize a modularly-invariant adaptation~\citep{shinoda2004} of the Parrinello-Rahman (PR) method~\citep{parrinello1981} for molecular dynamics (MD). This method was originally introduced to simulate the bulk properties of molecular systems in an isoenthalpic-isotension ensemble, including any phase transitions induced by the applied external stress~\citep{parrinello1981}. Such stress-controlled simulation methods adapted from molecular dynamics were previously implemented to study jamming~\citep{smith2014} and creep~\citep{srivastava2017} in granular packings, and recently to analyze flow-arrest transition in granular flows~\citep{srivastava2019}. However, this is the first study that utilizes these methods to simulate steady frictional granular flows under external shear stress in order to extract their constitutive rheological behavior. Simplified versions of such methods were also previously reported in simulations of non-equilibrium simple shear flows of Lennard-Jones fluids at a constant pressure and temperature to estimate their viscosity~\citep{evans1986,hood1987}, and recently for simulating the rheology of colloidal suspensions~\citep{wang2015}. The simulation framework described here can robustly simulate more complex flows beyond simple shear.

\begin{figure}
  \centerline{\includegraphics[width=0.95\columnwidth]{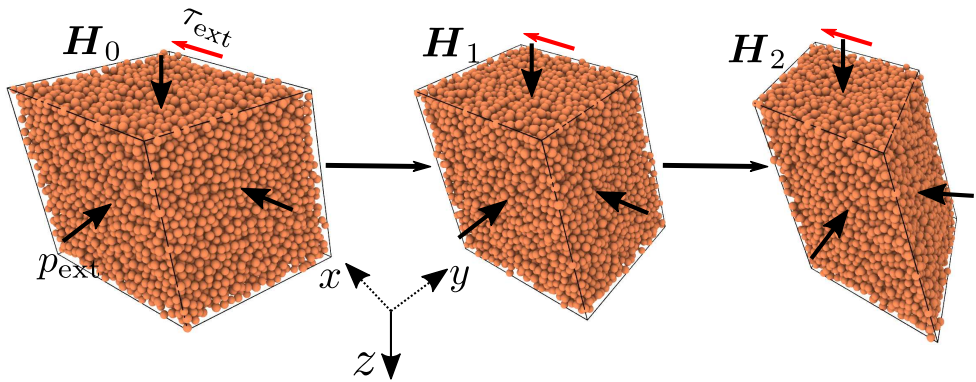}}
  \caption{Schematic of the simulation method: from left to right, the three images represent the configurations of a granular system at three consecutive simulation times during steady flow, while subjected to an external pressure $p_{\mathrm{ext}}$ and shear stress $\tau_{\mathrm{ext}}$. The triclinic periodic cell boundaries (in black) at three times are respectively represented by matrices $\boldsymbol{H}_{0}$, $\boldsymbol{H}_{1}$ and $\boldsymbol{H}_{2}$. The triclinic periodic cell volume is almost equal at all three times in steady flow. The dotted lines in the global coordinate system represent directions into the plane.}
\label{fig0}
\end{figure}

In the present simulations, a collection of particles contained within a 3D triclinic periodic cell is allowed to evolve under the application of a constant external stress tensor $\boldsymbol{\sigma}_{\mathrm{ext}}$, which is constrained by (i) $\left(1/3\right)\sigma_{\mathrm{ext},ii}=p_{\mathrm{ext}}$, (ii) $\sigma_{\mathrm{ext},ij}=\tau_{\mathrm{ext}}$ for $i,j=1,2$ and $2,1$, and (iii) $\sigma_{\mathrm{ext},ij}=0$ for all other Einstein indices $i\neq j$, as shown in the schematic in figure~\ref{fig0}. Because the traction at the boundaries of the periodic cell is prescribed, the periodic cell itself is allowed to dilate (or compact) and deform its shape in all possible ways, thus simulating the true bulk response of the granular material under external stress and pressure. The triclinic periodic cell is represented by a cell matrix $\mathsfbi{H}$ which is a concatenation of the three lattice cell vectors that define the periodicity of the system. The cell matrix is constrained to be upper-triangular and the internal stress tensor is symmetrized to prevent any spurious cell rotations, which was a problem in the original Parrinello-Rahman method. This was achieved differently using a positive-definite metric tensor in another variant of this method reported previously by~\citet{souza1997}. Upon the application of $\boldsymbol{\sigma}_{\mathrm{ext}}$, the equations of motion for $N$ particle positions and momenta $\{\boldsymbol{r}_{i},\boldsymbol{p}_{i}\}$, and the periodic cell matrix and its associated momentum tensor $\{\mathsfbi{H},\mathsfbi{P}_{g}\}$ are given by:
\begin{subeqnarray}
\dot{\boldsymbol{r}}_{i}&=&\frac{\boldsymbol{p}_{i}}{m_{i}}+\frac{\mathsfbi{P}_{g}}{W_{g}}\boldsymbol{r}_{i},\\[3pt]
\dot{\boldsymbol{p}}_{i}&=&\boldsymbol{f}_{i}-\frac{\mathsfbi{P}_{g}}{W_{g}}\boldsymbol{p}_{i}-\frac{1}{3N}\frac{\mathrm{Tr\left[\mathsfbi{P}_{g}\right]}}{W_{g}}\boldsymbol{p}_{i},\\[3pt]
\dot{\mathsfbi{H}}&=&\frac{\mathsfbi{P}_{g}}{W_{g}}\mathsfbi{H},\\[3pt]
\dot{\mathsfbi{P}_{g}}&=&V\left(\boldsymbol{\sigma}_{\mathrm{int}}-\boldsymbol{I}p_{\mathrm{ext}}\right)-\mathsfbi{H}\boldsymbol{\Sigma}\mathsfbi{H}^{T},
\label{eq5}
\end{subeqnarray}
where $\boldsymbol{f}_{i}$ is the net force on a particle $i$, $V$ is the variable volume of the periodic cell, $\boldsymbol{I}$ is the identity tensor, and $W_{g}$ is a `fictitious' mass associated with the inertia of the periodic cell. The stress quantities $\boldsymbol{\sigma}_{\mathrm{int}}$ and $\boldsymbol{\Sigma}$ are defined below. 

In the original PR method for molecular systems, the fictitious mass is suggested to be set as $W_{g}=Nk_{B}T/\omega_{g}^{2}$ for an efficient sampling of the isoenthalpic-isotension ensemble~\citep{martyna1996}. Here, $k_{B}$ is the Boltzmann constant, $T$ is intended temperature of the ensemble, and $\omega_{g}$ is the characteristic phonon frequency of the system. Such suggestions do not apply to the athermal flow simulations considered here. Analogously, the fictitious mass in the present case can be set as $W_{g}=Nk_{n}a^{2}/\omega_{g}^{2}$, where $k_{n}$ is the elastic constant associated with particle contacts (see Sec.~2.4), $a$ is the mean particle diameter, and $k_{n}a^{2}$ set the energy scale of system. The choice of $\omega_{g}$ controls the magnitude of stress fluctuations during steady granular flow, but it does not affect the rheology of flow within some upper and lower bounds of $\omega_{g}$, as was established by testing various values of $\omega_{g}$. A convenient value is $\omega_{g}=2.2\sqrt{m/k_{n}}$, where $m$ is the mean particle mass. Smaller values of $\omega_{g}$ resulted in larger stress fluctuations, whereas larger values of $\omega_{g}$ took longer simulation times to achieve steady flow. Similar analyses of the effect on $\omega_{g}$ on stress-controlled simulations were previously presented for non-equilibrium flow of Lennard-Jones fluids~\citep{evans1986,hood1987}. A comprehensive numerical analysis of the effect of $\omega_{g}$ on stress-controlled simulations of granular flows is a part of our ongoing work.

The first two terms of the right side of (\ref{eq5}\emph{d}) respectively represent the imbalance between bulk internal stress of the granular system $\boldsymbol{\sigma}_{\mathrm{int}}$ and external applied stress, which drives the motion of the periodic cell. The components of the bulk internal stress $\boldsymbol{\sigma}_{\mathrm{int}}$ are calculated as~\citep{walton1986,dacruz2005}:
\begin{equation}
\sigma_{\alpha\beta,\mathrm{int}}=\frac{1}{V}\sum_{i}\left[\sum_{j \neq i}\frac{1}{2}r_{\alpha,ij} f_{\beta,ij} + m_{i}v^{'}_{\alpha,i}v^{'}_{\beta,i}\right],
\label{eq6}
\end{equation}
where $\boldsymbol{r}_{ij}$ and $\boldsymbol{f}_{ij}$ are the branch vector and the force between two contacting particles $i$ and $j$. The fluctuating velocity $\boldsymbol{v}^{'}_{i}$ of particle $i$ is defined as the difference between velocity $\boldsymbol{v}_{i}$ of particle $i$ and mean shearing field velocity, such that $\boldsymbol{v}^{'}_{i}=\boldsymbol{v}_{i}-\left(\nabla \boldsymbol{v}\right)\boldsymbol{x}_{i}$, where $\nabla \boldsymbol{v}$ is bulk velocity gradient, and $\boldsymbol{x}_{i}$ is the position of particle $i$. Hereafter, the subscript `$\mathrm{int}$' will be dropped while referring to the internal state of the stress of the granular system. In (\ref{eq5}\emph{d}), the tensor $\boldsymbol{\Sigma}$ is defined as~\citep{shinoda2004}:
\begin{equation}
\boldsymbol{\Sigma}=\boldsymbol{H}_{0}^{-1}\left(\boldsymbol{\sigma}_{\mathrm{ext}}-\boldsymbol{I}p_{\mathrm{ext}}\right)\boldsymbol{H}_{0}^{T-1},
\label{eq7}
\end{equation}
where $\boldsymbol{H}_{0}$ is some reference state of the periodic cell, and $J^{-1}\mathsfbi{H}\boldsymbol{\Sigma}\mathsfbi{H}^{T}$ represents the `true' measure of the external deviatoric stress, which is defined with respect to the reference state~\citep{souza1997}. Here $J=\mathrm{det}\left[\boldsymbol{F}\right]$ is the Jacobian of the deformation gradient $\boldsymbol{F}$, which is defined as:
\begin{equation}
    \boldsymbol{F}=\boldsymbol{H}\boldsymbol{H}_{0}^{-1}.
    \label{eq7a}
\end{equation}

It is evident from (\ref{eq5}\emph{d}) that a difference between the internal stress and external applied stress drives the perpetual motion of the periodic cell during steady flow. In the case where the internal and external stress balance each other, the motion of the cell eventually stops because the external stress is not sufficient to continually drive the motion of the cell, thus enabling the precise identification of the yield stress of the granular system~\citep{srivastava2019}. As a result, this implementation of a constant external stress on the granular system prescribes the second Piola-Kirchoff measure of the external stress, or equivalently the thermodynamic tension~\citep{souza1997}. In the present simulations, the reference state is updated to the current state at the end of every time step of integration of the equations of motion, in order to minimize the deviation of internal strain energy from work done by the external stress. All the simulations are performed using the large-scale molecular dynamics software LAMMPS~\citep{plimpton1995}.

\subsection{Bulk Deformation}
Upon applying an external pressure $p_{\mathrm{ext}}$ and shear stress $\tau_{\mathrm{ext}}$ to a granular system, all the components of the macroscopic internal stress tensor $\boldsymbol{\sigma}$ evolve independently with time. Correspondingly, the triclinic periodic cell---represented by the matrix $\boldsymbol{H}$---also evolves with time from bulk volumetric and shear deformation. Figure~\ref{fig0} shows a schematic of the evolution of deformation of a triclinic periodic cell in steady flow as it is subjected to a constant external shear stress and pressure. The states of the triclinic periodic cell $\boldsymbol{H}$ are stored at every simulation time step (such as $\boldsymbol{H}_{0}$, $\boldsymbol{H}_{1}$ and $\boldsymbol{H}_{2}$ shown in figure~\ref{fig0}), and are used to compute the bulk velocity gradient in the periodic system, as described below.

Consider the position $\boldsymbol{r}(t)$ of a particle at a simulation time $t$ within the periodic cell, defined with respect to an origin (typically, one of the corners of the periodic cell). Its reduced coordinates $\boldsymbol{s}(t)$ can be defined by:
\begin{equation}
    \boldsymbol{r}(t)=\boldsymbol{H}(t)\boldsymbol{s}(t),
    \label{eq12a}
\end{equation}
such that $0\!<\!\boldsymbol{s}(t)\!<\!1$. The periodic tiling of the space by the triclinic cell $\boldsymbol{H}(t)$ implies that a spatial coordinate $\boldsymbol{r}^{'}(t)=\boldsymbol{H}(t)[\boldsymbol{s}(t)+\boldsymbol{\Delta}]$ represents the periodic image of $\boldsymbol{r}$, where $\boldsymbol{\Delta}$ is a vector of integers. The velocity $\boldsymbol{v}(t)=\dot{\boldsymbol{r}}(t)$ of the particle is defined such that:
\begin{equation}
    \boldsymbol{v}(t)=\dot{\boldsymbol{H}}(t)\boldsymbol{s}(t) + \boldsymbol{H}(t)\dot{\boldsymbol{s}}(t),
    \label{eq12b}
\end{equation}
where the first term represents the contribution from the bulk periodic cell deformation and the second term represents the fluctuating non-affine velocity. Consequently, a bulk velocity gradient can be defined as $\nabla \boldsymbol{v}(t)=\nabla_{\boldsymbol{r}}\left(\dot{\boldsymbol{H}}(t)\boldsymbol{s}(t)\right)$. Upon substituting (\ref{eq12a}):
\begin{equation}
    \nabla \boldsymbol{v}(t) = \dot{\boldsymbol{H}}(t)\boldsymbol{H}^{-1}(t).
    \label{eq12c}
\end{equation}
%The bulk velocity gradient can be further decomposed into a symmetric strain rate tensor $\boldsymbol{D}(t)=1/2\left(\nabla u(t) + \nabla u^{T} (t)\right)$ representing the rate of volumetric and shear deformation, and an anti-symmetric spin tensor $\boldsymbol{W}(t)=1/2\left(\nabla u(t) - \nabla u(t)^{T}\right)$ representing the rate of rigid-body rotation. 

\subsection{Contact Mechanics}
In the present simulations, frictional spherical particles interact only upon contact. The contact forces are modeled using a spring and a dashpot along with a \emph{static} yield criterion to model contact friction. This model was first developed by~\citet{cundall1979}, and since has been tested and implemented in various granular flow simulations~\citep{silbert2001,campbell2005,rycroft2009,sun2011}. Two contacting particles $\{i,j\}$ of diameters $\{a_{i},a_{j}\}$, masses $\{m_{i},m_{j}\}$, at positions $\{\boldsymbol{r}_i,\boldsymbol{r}_j\}$ with velocities $\{\boldsymbol{v}_i,\boldsymbol{v}_j\}$ and angular velocities $\{\boldsymbol{\omega}_i,\boldsymbol{\omega}_j\}$ are considered to be in contact if $\delta_{ij}=\frac{1}{2}(a_{i}+a_{j})-|\boldsymbol{r}_{ij}|>0$, where $\boldsymbol{r}_{ij}=\boldsymbol{r}_{i}-\boldsymbol{r}_{i}$ is the vector connecting their centroids; these quantities are tracked at every time step as they evolve from particle collisions or affine particle motion caused by triclinic cell deformation, as described in (\ref{eq5}\emph{a}). The contact normal force $\boldsymbol{f}_{nij}$ and tangential force $\boldsymbol{f}_{tij}$ on particle $i$ are given by:
\begin{subeqnarray}
\boldsymbol{f}_{nij}&=&k_{n} \delta_{ij} \boldsymbol{n}_{ij} - \gamma_{n} m_{e} \boldsymbol{v}_{nij},\\[3pt]
\boldsymbol{f}_{tij}&=&-k_{t} \boldsymbol{u}_{tij} - \gamma_{t} m_{e} \boldsymbol{v}_{tij},
\label{eq8}
\end{subeqnarray}
where $k_{n,t}$ and $\gamma_{n,t}$ are contact stiffness and damping constants, and $m_{e}=m_{i}m_{j}/(m_{i}+m_{j})$ is the effective mass. The corresponding force on particle $j$ is given by Newton's third law such that $\boldsymbol{f}_{ji}=\boldsymbol{f}_{ij}$. The unit normal along the axis of contact is given by $\boldsymbol{n}_{ij}=\boldsymbol{r}_{ij}/|\boldsymbol{r}_{ij}|$, and $\boldsymbol{v}_{nij}$ and $\boldsymbol{v}_{tij}$ are respectively the normal and tangential components of the relative velocity $\boldsymbol{v}_{ij}=\boldsymbol{v}_{i}-\boldsymbol{v}_{j}$ given by:
\begin{subeqnarray}
\boldsymbol{v}_{nij}&=&\left(\boldsymbol{v}_{ij}\boldsymbol{\cdot}\boldsymbol{n}_{ij}\right)\boldsymbol{n}_{ij},\\[3pt]
\boldsymbol{v}_{tij}&=&\boldsymbol{v}_{ij}-\boldsymbol{v}_{nij}-\tfrac{1}{2}\left(\boldsymbol{\omega}_{i}+\boldsymbol{\omega}_{j}\right)\times\boldsymbol{r}_{ij}.
\label{eq9}
\end{subeqnarray}
An elastic displacement $\boldsymbol{u}_{tij}$ representing shear in the tangential direction is tracked during the lifetime of a contact, and it evolves according to the following ODE:
\begin{equation}
\frac{\mathrm{d} \boldsymbol{u}_{tij}}{\mathrm{d} t}=\boldsymbol{v}_{tij}-\frac{\left(\boldsymbol{u}_{tij}\boldsymbol{\cdot}\boldsymbol{v}_{ij}\right)\boldsymbol{r}_{ij}}{|\boldsymbol{r}_{ij}|^{2}},
\label{eq10}
\end{equation}
with $\boldsymbol{u}_{tij}=0$ at the initiation of the contact.

Tangential friction between two contacting particles is modeled by a static yield criterion $|\boldsymbol{f}_{tij}|<\mu_{s}|\boldsymbol{f}_{nij}|$, which is always satisfied by limiting the tangential shear displacement $\boldsymbol{u}_{tij}$. The particle coefficient of sliding friction $\mu_{s}$ is a measure of its surface roughness, and significantly impacts the rheology of granular flow. The normal and tangential viscous damping at a contact are controlled by the coefficients of restitution $e_{n,t}=\mathrm{exp}(-\gamma_{n,t}t_{c}/2)$, where $t_{c}=\pi(2k_{n}/m_{e}-\gamma_{n}^{2}/4)^{-1/2}$ is the collision time between two contacting particles~\citep{silbert2001}. 

\subsection{Simulation Details}
The contact stiffness between particles $k_{n}$ and $k_{t}$ are set equal to each other. The normal damping constant $\gamma_{n}=0.5/t_{c}$ and the tangential damping constant $\gamma_{t}=0.5\gamma_{n}$. Initially, dilute configurations of granular systems at a solid volume fraction $\phi=0.05$ are subjected to a constant external shear stress and hydrostatic pressure. We simulate granular flow at three external pressures $p_{\mathrm{ext}}a/k_{n}=10^{-4},10^{-5},10^{-6}$, all in the limit of the rigid particle regime where the rheology is unaffected by the applied pressure and particle stiffness~\citep{dacruz2005,favierdecoulomb2017}. The external shear stress $\tau_{\mathrm{ext}}$ is varied from $\tau_{\mathrm{ext}}/p_{\mathrm{ext}}=0.0$ to $\tau_{\mathrm{ext}}/p_{\mathrm{ext}}=1.2$ to simulate flows at various shear rates, and three different realizations are simulated for each shear rate. Each simulation consists of $N\sim10^{4}$ frictional particles whose diameters are uniformly distributed between $0.9a$ and $1.1a$. Several particle coefficients of sliding friction ranging from $\mu_{s}=0.0$ to $\mu_{s}=0.3$ are analyzed to study the effect of friction on stress-controlled granular rheology. Contact mechanics between two particle is resolved by setting the simulation time step to $0.02t_{c}$. In the results presented below, time is scaled by $t_{c}$, length is scaled by $a$, energy is scaled by $k_{n}a^{2}$, and stress is scaled by $k_{n}/a$. 

\section{Evolution Towards Viscometric Flow}
\begin{figure}
  \centerline{\includegraphics[width=0.9\columnwidth]{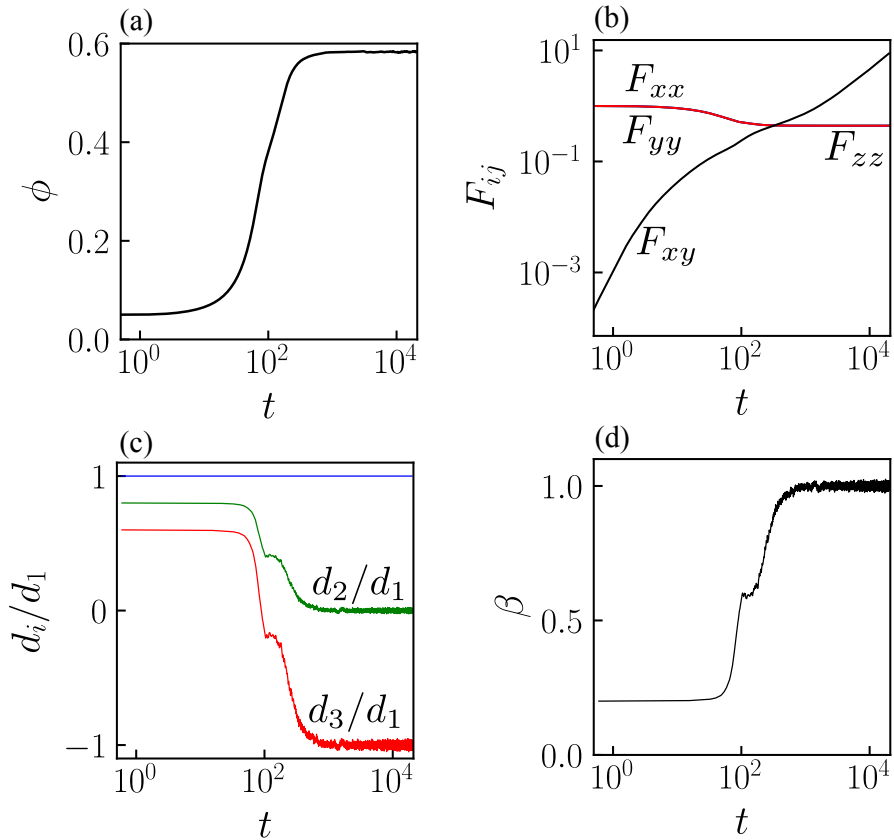}}
  \caption{Evolution with time $t$ of (a) solid volume fraction $\phi$, (b) components of the deformation gradient tensor $F_{ij}$, (c) $d_{i}/d_1$, where $d_{i}$ are the eigenvalues of $\boldsymbol{D}$, and (d) vorticity parameter $\beta$ for a particular case of interparticle friction $\mu_{s}=0.3$, applied pressure $p_{\mathrm{ext}}=10^{-4}$, and applied shear stress  $\tau_{\mathrm{ext}}=5\times10^{-5}$.}
\label{fig1}
\end{figure}
When the external shear stress $\tau_{\mathrm{ext}}$ and pressure $p_{\mathrm{ext}}$ are switched on at $t\!=\!0$, a dilute assembly of particles at an initial solid volume fraction $\phi\!=\!0.05$ responds with rapid volumetric compaction, as shown by the evolution of $\phi$ in figure~\ref{fig1}(\emph{a}) for a particular case of interparticle friction $\mu_{s}=0.3$, $p_{\mathrm{ext}}=10^{-4}$ and $\tau_{\mathrm{ext}}=5\times10^{-5}$. To estimate the total deformation accumulated by the material beyond isotropic compaction, we calculate the deformation gradient $\boldsymbol{F}(t)=\boldsymbol{H}(t)\boldsymbol{H}_{0}^{-1}$ as defined in (\ref{eq7a}), where $\boldsymbol{H}_{0}$ is the periodic cell at $t\!=\!0$. The rapid volumetric compaction at early times is seen by an equivalent decrease in $F_{ii}$ in figure~\ref{fig1}(\emph{b}) for $i=x,y,z$. The shear component $F_{xy}$ exhibits a super-linear increase at early times as a result shear deformation at low solid volume fractions in the absence of any significant resistance to the applied shear. At long times, $F_{xy}$ increases linearly with time, while $F_{ii}$ is constant and the other two shear components are negligible, thus indicating the achievement of steady incompressible viscometric flow, i.e., $\boldsymbol{F}(t)=\boldsymbol{I}+t\boldsymbol{M}$~\citep{coleman2012viscometric}, where $\boldsymbol{M}$ is a constant tensor, and which is a special case of motion with constant stretch history~\citep{noll1962} where the deviatoric stress depends on the form of $\boldsymbol{M}$~\citep{coleman2012viscometric}. Several important and well-studied flows such as Couette flow, Poiseuille flow, simple shearing flow and some specific cases of torsional flows can be categorized as viscometric flows~\citep{coleman2012viscometric}. Although our focus here is on steady flows, the simulation method and rheological analysis described above provide the capability to calibrate a general history-dependent rheological model defined in (\ref{eq1}) for transient granular flows under constant or time-varying applied stresses. 

% The rapid increase in $\phi$ is accompanied by a rapid increase in the number of interparticle contacts, as shown by the evolution of coordination number $Z_{2}$ in Figure~\ref{fig1}(\emph{b}). Typically the coordination number $Z$ is defined as the ratio of total number of interparticle contacts $N_c$ and the number of particles $N$. However, it is customary to ignore rattler particles, i.e., particles with lesser than two contacts while computing the coordination number, because these particles are not involved in the load-bearing particle contact network~\citep{sun2011}. Here, $Z_{2}$ denotes the rattler-free coordination number. After the initial transients, the system transitions into steady flow at long times, which is characterized by fluctuating $\phi$ and $Z_{2}$ around constant mean values.

Further insight into the nature of viscometric flow is given by the eigenvalue decomposition of the symmetric tensor $\boldsymbol{D}(t)$. We use the convention that the three orthonormal eigenvectors of $\boldsymbol{D}$: $\boldsymbol{\hat{d}}_{1}$, $\boldsymbol{\hat{d}}_{2}$ and $\boldsymbol{\hat{d}}_{3}$ are ordered in the decreasing order of signed eigenvalues $d_1$, $d_2$ and $d_3$. Figure~\ref{fig1}(\emph{c}) shows the evolution of $d_2/d_1$ and $d_3/d_1$ as a function of time. At early times, the sum of eigenvalues is positive, which corresponds with rapid volumetric compaction as described above. The long time steady state flow is characterized by $d_3=-d_1$ and $d_2=0$, which is a signature of planar flow, where the flow plane is spanned by $\boldsymbol{\hat{d}}_{1}$ and $\boldsymbol{\hat{d}}_{3}$. To further ascertain the nature of planar flow, we calculate a vorticity parameter $\beta$ defined as: 
\begin{equation}
    \beta = \frac{1}{\dot{\gamma}}\frac{\boldsymbol{W}:\boldsymbol{G}}{\boldsymbol{G}:\boldsymbol{G}},
    \label{eq13}
\end{equation}
where $\boldsymbol{G}=\boldsymbol{\hat{d}}_{3}\boldsymbol{\hat{d}}_{1}-\boldsymbol{\hat{d}}_{1}\boldsymbol{\hat{d}}_{3}$. Figure~\ref{fig1}(\emph{d}) shows the evolution of $\beta$ with time. When the system transition into steady state flow at long times, $\beta=1$, indicating simple shear deformation in the flow plane, thus confirming the viscometric nature of flow. During the transient evolution at early times, $0<\beta<1$, indicating a complex flow behavior that is a mix of vorticity-free elongational flow ($\beta=0$) and simple shear flow ($\beta=1$)~\citep{wagner2016,giusteri2018}. However, the flow is homogeneous at all times within the periodic cell during steady state, with no spatial gradients of the local strain rate. 

We emphasize that the steady homogeneous shear flow states achieved in the present simulations considerably simplify the rheological model in (\ref{eq1}). Because the eigenvectors of $\boldsymbol{D}$ are uniform in space and time, the material derivative $\dot{\boldsymbol{D}}=0$, and the local material rotation is equivalent to flow vorticity. In this particular case of steady homogeneous flow with constant stretch history, the rheology is equally well-represented by the following form of (\ref{eq1})~\citep{larson1985flows,brunn2003,giusteri2018}:
\begin{eqnarray}
    \boldsymbol{\sigma} &=& p\boldsymbol{I}+\eta_{1}\boldsymbol{D}+\eta_{2}\left[\boldsymbol{D}^{2}-\frac{\mathrm{tr}\left(\boldsymbol{D}^{2}\right)}{3}\boldsymbol{I}\right]+\eta_{3}\left[\boldsymbol{D}\boldsymbol{W}-\boldsymbol{W}\boldsymbol{D}\right]\nonumber \\
    &&+\kappa_{1}\frac{\boldsymbol{D}}{|\boldsymbol{D}|}+\kappa_{2}\left[\frac{\boldsymbol{D}^{2}}{|\boldsymbol{D}|^{2}}-\frac{\mathrm{tr}\left(\boldsymbol{D}^{2}\right)}{3|\boldsymbol{D}|^{2}}\boldsymbol{I}\right].
    \label{eq1e}
\end{eqnarray}
We emphasize that in unsteady or inhomogeneous flows where the material rate of rotation can differ from flow vorticity, several criteria for classifying local flow kinematics have been prescribed~\citep{schunk1990,thompson2005}, and they can be incorporated in the current rheological model. 

In steady state, the bulk rheological quantities fluctuate around their mean values, as seen in figure~\ref{fig1}(\emph{a}-\emph{d}). In order to achieve robust statistics, every simulation is run for at least $10^{7}$ time steps to guarantee the achievement of steady state flow. This is especially necessary near the critical yield stress, where steady state equilibration takes a long time. Upon achieving steady state, each simulation is continued to run for at least another $10^{6}$ time steps, during which the all data of interest are recorded at every $10$ time steps and averaged to estimate their steady mean value. The statistical uncertainty associated with mean estimation is measured by its standard error using a block averaging method~\citep{flyvbjerg1989}. This method not only provides robust estimates of uncertainty around a mean value, but also indicates if the data has any long-time correlations, which would necessitate longer simulation runs for meaningful averaging.

\section{Model Calibration}\label{evol}

In this section, friction-dependent functional forms of all flow coefficients in (\ref{eq1e}) will be described. The material constants associated with these flow coefficients are extracted from the DEM simulation data by utilizing the fact that the four tensors $\boldsymbol{I}$, $\boldsymbol{D}$, $\left(\boldsymbol{D}^{2}\!-\!\frac{\mathrm{tr}\left(\boldsymbol{D}^{2}\right)}{3}\boldsymbol{I}\right)$ and $\left(\boldsymbol{D}\boldsymbol{W}-\boldsymbol{W}\boldsymbol{D}\right)$ are orthogonal to each other in viscometric flows.

\subsection{Flow Functions: $\eta_{1}$ and $\kappa_{1}$}
The two flow coefficients $\eta_{1}$ and $\kappa_{1}$ have a first-order contribution (in terms of $\boldsymbol{D}$) to the total stress $\boldsymbol{\sigma}$, and they provide a measure of the shear stress in viscometric flow. These coefficients are estimated by:
\begin{equation}
\eta_{1}\dot{\gamma}+\kappa_{1} = \frac{1}{2\dot{\gamma}}\boldsymbol{\sigma}:\boldsymbol{D},
    \label{eq14}
\end{equation}
where $\tau\!=\!\frac{1}{2\dot{\gamma}}\boldsymbol{\sigma}\!:\!\boldsymbol{D}$ is the total flow-induced shear stress in the system.

\begin{figure}
  \centerline{\includegraphics[width=0.95\columnwidth]{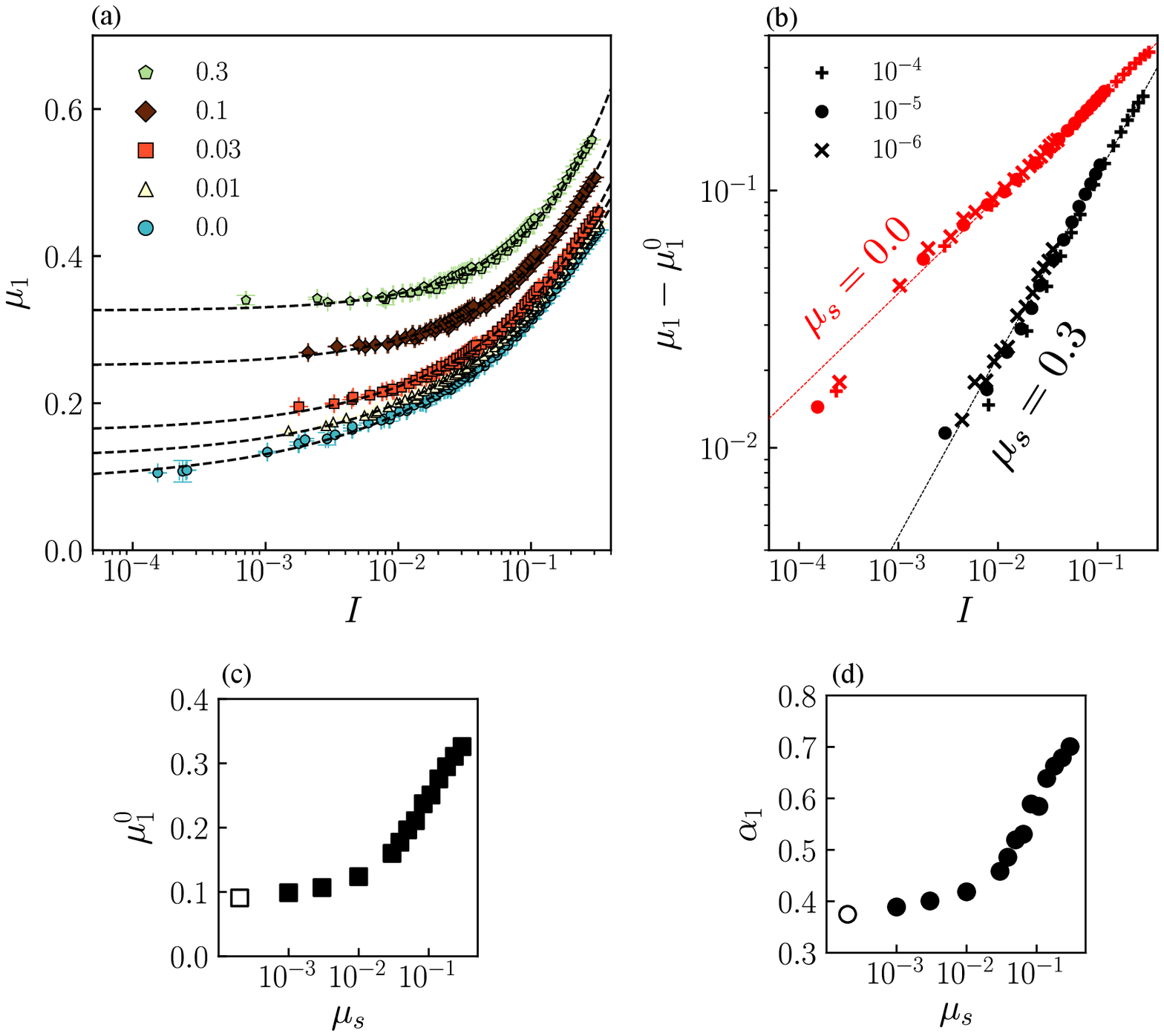}}
  \caption{(a) Stress ratio $\mu_{1}$ as a function of inertial number $I$ for five interparticle friction $\mu_{s}$ (see legend) at three applied pressures: $p_{\mathrm{ext}}=10^{-4},10^{-5},10^{-6}$. The vertical and horizontal error bars represent the standard error in the calculation of $\mu_1$ and $I$ respectively. The black dashed lines represent fits for each $\mu_{s}$ given in (\ref{eq15}). (b) Variation of $\mu_{1}-\mu_{1}^{0}$ with $I$ at three applied pressures (see legend) for particles with $\mu_{s}=0.0$ (red) and $\mu_{s}=0.3$ (black). The dotted lines represent power-law fits from (\ref{eq15}). (c) Variation of $\mu_{1}^{0}$ and (d) $\alpha_{1}$ with $\mu_{s}$. The open symbols in (c) and (d) indicate the values for zero friction.}
\label{fig2}
\end{figure}

Previous research has shown that shear flow of granular materials can be well-described by a local rheological relationship between a stress ratio $\mu$ and an inertial number $I$~\citep{jop2006}. In the present model, the stress ratio (hereby written with a subscript $1$) is $\mu_{1}\!=\!(\eta_{1}\dot{\gamma}\!+\!\kappa_{1})/p$, where $\eta_{1}\dot{\gamma}/p$ is the rate-dependent contribution and $\kappa_{1}/p$ is the rate-independent contribution. As such, $\eta_{1}$ represents the effective shear viscosity and $\kappa_{1}/p$ represents the yield coefficient as $I\to0$, which is associated with a critical volume fraction discussed in Sec.~4.4. Figure~\ref{fig2}(\emph{a}) shows the variation of $\mu_{1}$ with $I$ for five interparticle friction $\mu_{s}$ at three $p_{\mathrm{ext}}$. All the curves at various pressures collapse onto a master curve for each $\mu_{s}$, which can be approximated by a power law for dense granular flows described in several previous studies~\citep{degiuli2015,degiuli2016,favierdecoulomb2017,salerno2018}:
\begin{equation}
\mu_{1}=\mu_{1}^{0}+A_{1}I^{\alpha_1},
    \label{eq15}
\end{equation}
where, $\mu_{1}^{0}$, $A_1$ and $\alpha_1$ are fitting parameters. In the quasi-static, rate-independent regime where $I\!\to\!0$, the stress ratio reaches a constant value $\mu_{1}\!\to\!\mu_{1}^{0}$, which is equivalent to $\kappa_{1}/p$ in the rheological model. In this regime shear stress saturates towards a threshold value, while the pressure is well-controlled at its prescribed value, thus indicating the approach towards a yield stress. Although the rheology is unaffected by the applied pressure, as also observed in~\citep{favierdecoulomb2017}, lower values of inertial numbers are achieved when the confining pressure is low, as seen by the variation of $\mu_{1}-\mu_{1}^{0}$ with $I$ for three pressures and two $\mu_{s}$ in figure~\ref{fig2}(\emph{b}). Previous pressure and shear rate controlled simulations had demonstrated that the transition from quasi-static to inertial flow regimes occurs at lower inertial number for lower confining pressure~\citep{favierdecoulomb2017}, thus confirming the current observations. However, the present simulations produce highly stochastic flows in the quasi-static regime, which often arrest in the vicinity of the static yield coefficient~\citep{srivastava2019}. Therefore, the rheology at low inertial numbers is not well-resolved for low pressures, especially for intermediate interparticle friction, as seen in figure~\ref{fig2}(\emph{a}). Recent experiments~\citep{perrin2019} and simulations~\citep{degiuli2017} have indicated that the local rheology of frictional granular materials possibly exhibits hysteresis at very low inertial numbers, which would also prohibit very slow flows in the present stress-controlled simulations.

The quasi-static stress ratio increases with friction from $\mu_{1}^{0}\!=\!0.09$ for frictionless particles to $\mu_{1}^{0}\!=\!0.33$ for particles with high friction, as shown in figure~\ref{fig2}(\emph{c}). The non-zero value of $\mu_{1}^{0}$ for frictionless particles is consistent with previous simulations~\citep{peyneau2008} and experiments~\citep{clavaud2017,perrin2019} that demonstrated a non-zero internal friction angle for frictionless granular material. The value of $\mu_{1}^{0}$ at high friction is consistent with previous simulations and experiments~\citep{boyer2011a,salerno2018,srivastava2019}, and is also similar to the critical stress ratio from the critical state theory~\citep{schofield1968critical}. The power-law exponent varies monotonically between $\alpha_1\!=\!0.37$ for frictionless particles to $\alpha_1\!=\!0.7$ for particles with high friction, as seen in figure~\ref{fig2}(\emph{d}). Although the exponent for frictionless particles correspond well with prior theoretical predictions~\citep{degiuli2015,degiuli2016}, the exponent at high friction is smaller than theoretical predictions of $\alpha_1=1.0$~\citep{degiuli2015,degiuli2016}. This could be attributed to a lack of data at low inertial numbers and the associated sensitivity of power-law fitting.

% The flow coefficient $\eta_{1}$, which corresponds to a shear viscosity, can be re-scaled to its dimensionless value as $\frac{p^{-1/2}\eta_{1}}{d\sqrt{\rho}}$. Figure~\ref{fig2}(\emph{d}) shows the two friction-dependent limits of viscosity $\eta_{1}^{0}$ and $\eta_{1}^{\infty}$ corresponding to quasi-static regime and high shear rate flows respectively. The low shear rate viscosity $\eta_{1}^{0}$ decreases with increasing $\mu_{s}$, whereas the high shear rate viscosity $\eta_{1}^{\infty}$ is independent of $\mu_{s}$. At low shear rates, the average lifetime of a contact between two particles is long enough that friction crucially dominates the rheology of a granular system. However, at high shear rates, the average lifetime of a contact is short, and the effective shear viscosity is independent of friction.

\subsection{Flow Functions: $\eta_{2}$ and $\kappa_{2}$}
\begin{figure}
  \centerline{\includegraphics[width=0.95\columnwidth]{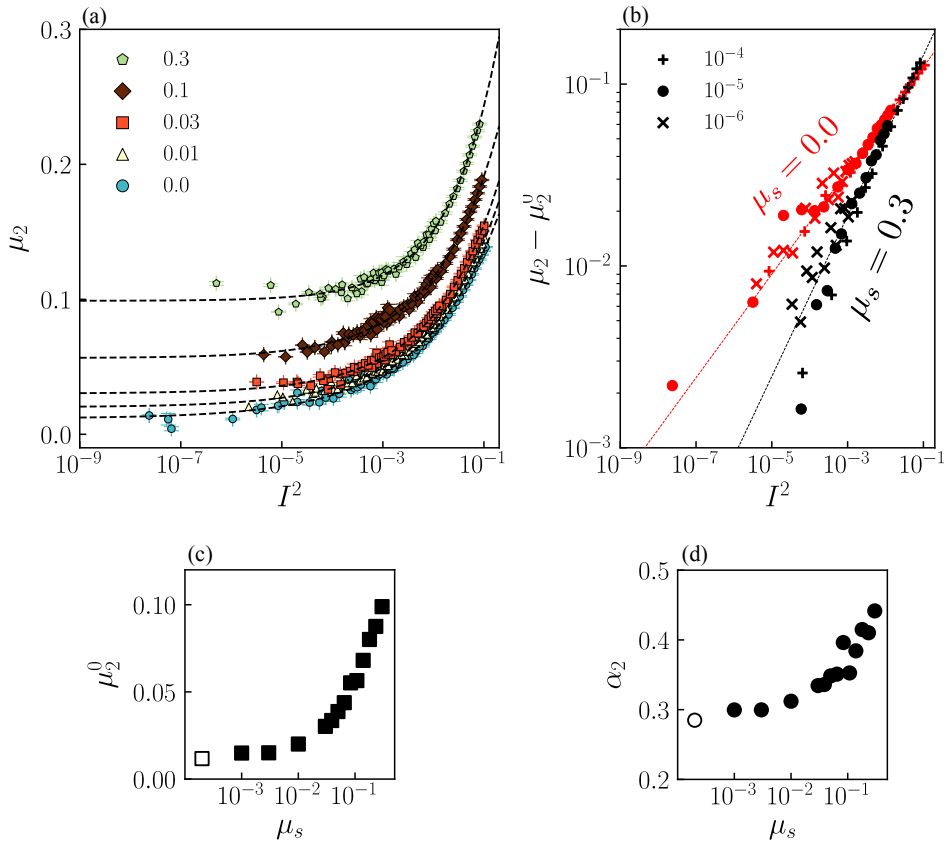}}
  \caption{(a) Second stress ratio $\mu_{2}$ as a function of inertial number $I^2$ for five interparticle friction $\mu_{s}$ (see legend) at three applied pressures: $p_{\mathrm{ext}}=10^{-4},10^{-5},10^{-6}$. The vertical and horizontal error bars represent the standard error in the calculation of $\mu_2$ and $I^2$ respectively. The black dashed lines represent fits for each $\mu_{s}$ given in (\ref{eq17}). (b) Variation of $\mu_{2}-\mu_{2}^{0}$ with $I^2$ at three applied pressures (see legend) for particles with $\mu_{s}=0.0$ (red) and $\mu_{s}=0.3$ (black). The dotted lines represent power-law fits from (\ref{eq17}). (c) Variation of $\mu_{2}^{0}$ and (d) $\alpha_{2}$ with $\mu_{s}$. The open symbols in (c) and (d) indicate the values for zero friction.}
\label{fig3}
\end{figure}
In addition to the shear stress contribution to the total internal stress, there are non-negligible second-order contributions that are typically observed in the flow of non-Newtonian fluids. In a viscometric description of such fluids, these effects are characterized by normal stress difference functions~\citep{guazzelli2018}. In the present rheological model, $\eta_{2}$ and $\kappa_{2}$ represent one set of such rate-dependent and rate-independent contributions. These coefficients are estimated by:
\begin{equation}
\eta_{2}\dot{\gamma}^{2}+\kappa_{2} = \frac{3}{2\dot{\gamma}^{2}}\boldsymbol{\sigma}:\left(\boldsymbol{D}^{2}-\frac{\mathrm{tr}\left(\boldsymbol{D}^{2}\right)}{3}\boldsymbol{I}\right),
    \label{eq16}
\end{equation}
and they represent the difference between mean normal stress in the flow plane and normal stress in the vorticity direction. A second stress ratio similar to $\mu_{1}$ is defined as $\mu_{2}\!=\!\left(\eta_{2}\dot{\gamma}^{2}\!+\!\kappa_{2}\right)/p$, where $\eta_{2}\dot{\gamma}^{2}/p$ is the rate-dependent contribution and $\kappa_{2}/p$ is the rate-independent contribution. As such, $\eta_{2}$ represents a normal viscosity and $\kappa_{2}/p$ represents the threshold value as $I\to0$. Figure~\ref{fig3}(\emph{a}) shows the variation of $\mu_{2}$ with the square of inertial number $I$ for five $\mu_{s}$ and three $p_{\mathrm{ext}}$. All the curves at various pressures collapse onto a master curve for each $\mu_{s}$, which can be approximated by a power law:
\begin{equation}
\mu_{2}=\mu_{2}^{0}+A_{2}\left(I^{2}\right)^{\alpha_2},
    \label{eq17}
\end{equation}
where, $\mu_{2}^{0}$, $A_2$ and $\alpha_2$ are fitting parameters. In a manner similar to $\mu_{1}$, the quasi-static values of $\mu_{2}$ at low inertial numbers are accessed for low confining pressures, as shown by the variation of $\mu_{2}-\mu_{2}^{0}$ with $I^{2}$ in figure~\ref{fig3}(\emph{b}) for two different $\mu_{s}$ that appear to collapse onto a single curve. However, the data at low inertial numbers is also noisy, resulting from the stochastic nature of slow granular flows, and increased noise in the measured data.

In the quasi-static regime, $\mu_{2}$ tends towards a constant value $\mu_{2}\!\to\!\mu_{2}^{0}$, which is equivalent to $\kappa_{2}/p$ in the rheological model. Its value varies monotonically from $\mu_{2}^{0}\!=\!0.01$ for frictionless particles to $\mu_{2}^{0}\!=\!0.1$ for particles with high friction, as shown in figure~\ref{fig3}(\emph{c}). The non-zero value of $\mu_{2}^{0}$ for particles with high friction indicates that normal stress effects are present even in the quasi-static regime of flow, thus indicating a mild anisotropic nature of the yield surface that is commonly assumed to be isotropic (in the Drucker-Prager sense) within the $\mu(I)$ rheology~\citep{jop2006}, but has been shown to be anisotropic in recent simulations~\citep{thornton2010,li2012}. The power-law exponent varies monotonically between $\alpha_2=0.28$ for frictionless particles to $\alpha_2=0.44$ for particles with high friction, as shown in figure~\ref{fig3}(\emph{d}).

% Analogous to the shear viscosity $\eta_{1}$, a normal viscosity $\eta_{2}$ is also extracted from the simulation data, which can be re-scaled to its dimensionless form: $\eta_{2}/d^{2}\rho$. Unlike $\eta_{1}$, $\eta_{2}$ is independent of the applied pressure. Figure~\ref{fig4}(\emph{d}) shows the two friction-dependent limits of dimensionless normal viscosity $\eta_{2}^{0}$ and $\eta_{2}^{\infty}$ corresponding to quasi-static and at high shear rate flows. The quasi-static viscosity is always larger in magnitude to the high rate viscosity for $\mu_{s}$. While $\eta_{2}^{0}$ is about three times larger for frictionless particles than particles with high friction, $\eta_{2}^{\infty}$ is largely independent of $\mu_{s}$.

\subsection{Flow Function: $\eta_{3}$}
\begin{figure}
  \centerline{\includegraphics[width=0.95\columnwidth]{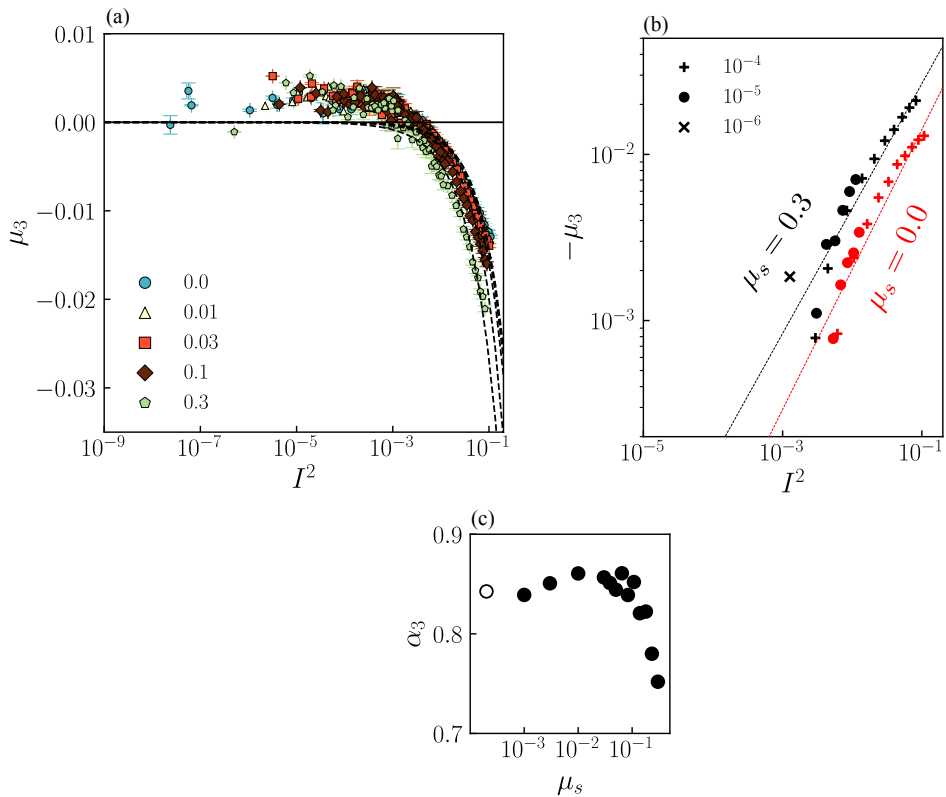}}
  \caption{(a) Third stress ratio $\mu_{3}$ as a function of inertial number $I^2$ for five interparticle friction $\mu_{s}$ (see legend) at three applied pressures: $p_{\mathrm{ext}}=10^{-4},10^{-5},10^{-6}$. The vertical and horizontal error bars represent the standard error in the calculation of $\mu_3$ and $I^2$ respectively. The black dashed lines represent fits for each $\mu_{s}$ given in (\ref{eq19}). (b) Variation of $-\mu_{3}$ with $I^2$ at three applied pressures (see legend) for particles with $\mu_{s}=0.0$ (red) and $\mu_{s}=0.3$ (black). The dotted lines represent power-law fits from (\ref{eq19}). (c) Variation of $\alpha_{3}$ with $\mu_{s}$. The open symbol in (c) indicates the values for zero friction.}
\label{fig4}
\end{figure}

An additional second-order contribution to the total stress emerges through the rate-dependent flow coefficient $\eta_{3}$, which is estimated by:
\begin{equation}
\eta_{3}\dot{\gamma}^{2} = \frac{1}{8\dot{\gamma}^{2}}\boldsymbol{\sigma}:\left(\boldsymbol{D}\boldsymbol{W}-\boldsymbol{W}\boldsymbol{D}\right),
    \label{eq18}
\end{equation}
and for viscometric flows, it represents the difference between the two normal stresses in the flow plane. A third stress ratio $\mu_{3}$ is defined as: $\mu_{3}\!=\!\eta_{3}\dot{\gamma}^{2}/p$. where $\eta_{3}$ represents an additional normal viscosity. Figure~\ref{fig4}(\emph{a}) shows the variation of $\mu_{3}$ as a decreasing function of $I^{2}$ for five $\mu_{s}$ at three $p_{\mathrm{ext}}$. All the curves at various pressures collapse onto a master curve for each $\mu_{s}$, which can be approximated by the following power law:
\begin{equation}
\mu_{3}=-A_{3}\left(I^{2}\right)^{\alpha_3},
    \label{eq19}
\end{equation}
where $A_{3}$ and $\alpha_3$ are fitting parameters.

The stress ratio $\mu_{3}$ exhibits a transition from negative values at high inertial numbers to small positive values in the quasi-static regime for all $\mu_{s}$, as seen in figure~\ref{fig4}(\emph{a}). Although the small positive value of $\mu_{3}$ in the quasi-static regime is intriguing, its existence is debated (see Section 5.2 below) and this effect is not included in our rheological model. As such, a simple power law in (\ref{eq19}) well-predicts the variation of $\mu_{3}$ with $I$, as also seen by the variation of $-\mu_{3}$ with $I^{2}$ in figure~\ref{fig4}(\emph{b}). The power-law exponent $\alpha_3$ varies slightly between $0.85$ and $0.75$ from low to high friction, as shown in figure~\ref{fig4}(\emph{c}).

For steady homogeneous simple shear flows simulated in this work, the constitutive model for viscometric flows in (\ref{eq1e}) reduces to the following relationships between the components of symmetric stress tensor $\boldsymbol{\sigma}$ and the three stress ratios, in the case of a shear flow along the $x$ direction and flow gradient along the $y$ direction:
\begin{subeqnarray}
\sigma_{xx}&=&p\left(1+\frac{\mu_{2}}{3}-2\mu_{3}\right),\\[3pt]
\sigma_{yy}&=&p\left(1+\frac{\mu_{2}}{3}+2\mu_{3}\right),\\[3pt]
\sigma_{zz}&=&p\left(1-\frac{2\mu_{2}}{3}\right),\\[3pt]
\sigma_{xy}&=&p\mu_{1},\\[3pt]
\sigma_{yz}&=&\sigma_{xz}=0.
\label{eq1f}
\end{subeqnarray}

% The rate-dependent flow coefficient characterized by the normal viscosity $\eta_{3}$ can be re-scaled to its dimensionless form: $\eta_{3}/d^{2}\rho$. The two limits of (negative) dimensionless viscosity: $\eta_{3}^{0}$ in the quasi-static regime $\eta_{3}^{0}$, and $\eta_{3}^{\infty}$ at high shear rates is displayed as a function of $\mu_{s}$ in figure~\ref{fig6}(\emph{d}). The quasi-static viscosity $\eta_{3}^{0}$ is approximately an order of magnitude larger than $\eta_{3}^{\infty}$, and they both are about thrice as large for high friction particles compared to frictionless particles.

\subsection{Granular Flow-Induced Dilation}
\begin{figure}
  \centerline{\includegraphics[width=0.95\columnwidth]{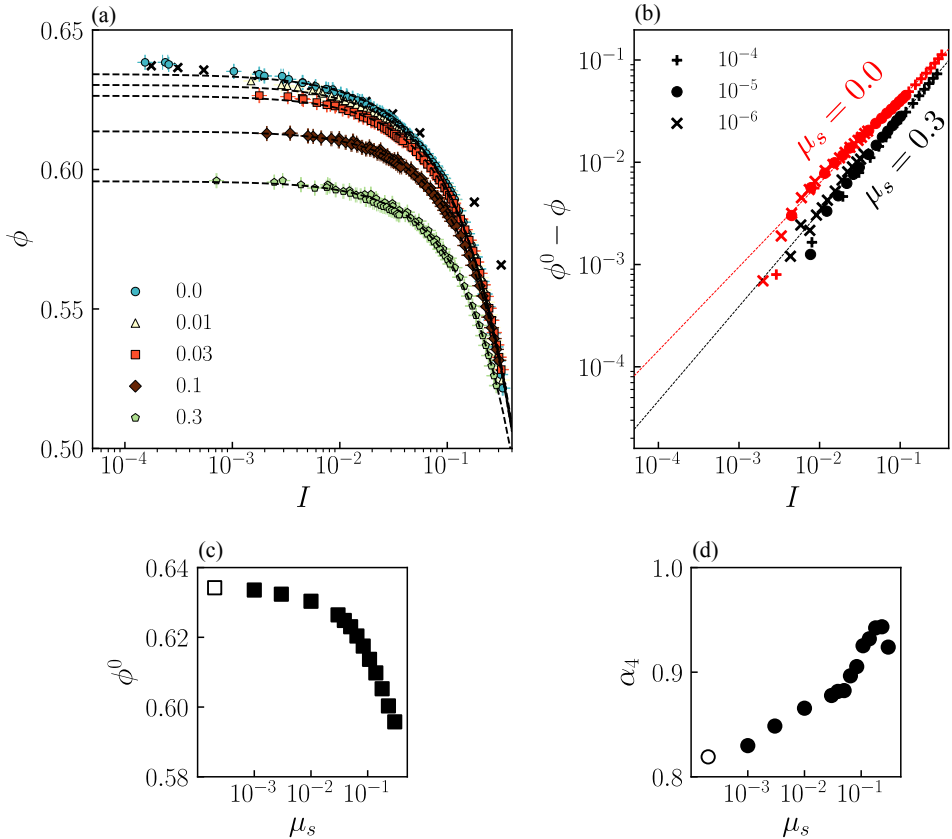}}
  \caption{(a) Solid volume fraction $\phi$ as a function of inertial number $I$ for five interparticle friction $\mu_{s}$ (see legend) at three applied pressures: $p_{\mathrm{ext}}=10^{-4},10^{-5},10^{-6}$. The vertical and horizontal error bars represent the standard error in the calculation of $\phi$ and $I$ respectively. The black dashed lines represent fits for each $\mu_{s}$ given in (\ref{eq20}). The black crosses represent the data from~\citet{peyneau2008}. b) Variation of $\phi^{0}-\phi$ with $I$ at three applied pressures (see legend) for particles with $\mu_{s}=0.0$ (red) and $\mu_{s}=0.3$ (black). The dotted lines represent power-law fits from (\ref{eq20}). (c) Variation of $\phi^{0}$ and (d) $\alpha_{4}$ with $\mu_{s}$. The open symbols in (c) and (d) indicate the values for zero friction.}
\label{fig5}
\end{figure}

The solid volume fraction $\phi$ of granular materials is highly sensitive to pressure and the rate of shear flow. These materials compact (jam) under the action of external pressure. However, under the action of external shear stress they dilate in order to flow, and the extent of dilation is higher for faster flows. In the present simulations, $\phi$ is not prescribed, and the system is allowed to freely attain its steady state solid volume fraction in response to the external stress and pressure. As such, we extract a dilatancy law relating the steady-state $\phi$ with the inertial number $I$ of the flow. Figure~\ref{fig5}(a) shows the variation of $\phi$ with $I$ for five $\mu_{s}$ at three $p_{\mathrm{ext}}$. All the curves at various pressures collapse onto a master curve for each $\mu_{s}$, which can be approximated by a power law as described in several previous studies~\citep{degiuli2015,degiuli2016,favierdecoulomb2017}:
\begin{equation}
\phi=\phi^{0}-A_{4}I^{\alpha_4},
    \label{eq20}
\end{equation}
$\phi^{0}$, $A_{4}$ and $\alpha_4$ are fitting parameters. The applied pressure moderately affects volume fraction $\phi$, with lower $\phi$ at lower pressures, as seen in figure~\ref{fig5}(b). It has been previously demonstrated that for sufficiently rigid particles (or equivalently, low enough applied pressures) in the hard particle limit, the effect of pressure is negligible on the volume fraction of granular material at onset of flow~\citep{favierdecoulomb2017}.

The quasi-static solid volume fraction $\phi^{0}$ varies significantly with $\mu_{s}$ ranging from $\phi^{0}\!=\!0.63$ for frictionless particles to $\phi^{0}\!=\!0.59$ for particles with high friction, as shown in figure~\ref{fig5}(c). Such a dependence of $\phi^{0}$ on friction has been previously demonstrated in 2D~\citep{dacruz2005} and 3D simulations~\citep{sun2011}, and confirmed in recent experiments~\citep{tapia2019}. The similarity between $\phi^{0}\!=\!0.63$ for frictionless particles and the solid volume fraction of random close packing of mono-disperse spheres indicates that frictionless particles do not dilate at the onset of flow, which is consistent with prior simulations~\citep{peyneau2008} and experiments~\citep{clavaud2017}. For particles with high friction, $\phi^{0}$ is consistent with the critical state solid volume fraction from the critical state theory~\citep{schofield1968critical}, and previous simulations~\citep{sun2011,favierdecoulomb2017,srivastava2019} and experiments~\citep{boyer2011a,tapia2019}.

The power-law exponent varies between $\alpha_4=0.82$ for frictionless particles and $\alpha_4=0.92$ for high friction particles, as shown in figure~\ref{fig5}(d). The value of this exponent at high friction is similar to previous theoretical predictions of a unity exponent~\citep{degiuli2015,degiuli2016}. For frictionless particles, our prediction of $\alpha_4$ does not correspond well with theoretical prediction~\citep{degiuli2015,degiuli2016} of $\alpha_4=0.35$, and a previous simulation study~\citep{peyneau2008} that demonstrated $\alpha_4=0.39$. However, as shown in figure~\ref{fig5}(a), our data corresponds well with the simulations of ~\citet{peyneau2008} at low and moderate inertial numbers, but deviates slightly at higher inertial numbers, resulting in large changes to the power-law exponent.

\section{Normal Stress Differences and their Microstructural Origins}
The non-negligible second-order contributions to stress in viscometric granular flows indicate the presence of normal stress differences. Previous research on sheared granular and suspension flows has demonstrated the existence of normal stress differences~\citep{silbert2001,alam2005,rycroft2009,sun2011,couturier2011,boyer2011,weinhart2013,saha2016,seto2018,guazzelli2018}, and these differences have been attributed to flow-induced fluctuating velocity effects in dilute granular flows~\citep{saha2016} and microstructural effects in dense suspension flows~\citep{seto2018}. Particularly, normal stress differences can arise either from: (1) a misalignment of $\boldsymbol{\sigma}$ and $\boldsymbol{D}$ in the flow plane, known as the first normal stress difference, or (2) from the anisotropy of normal stress between the flow plane and the vorticity direction, known as the second normal stress difference. 

In this section, we describe normal stress differences and their microstructural origins in dense viscometric granular flows. The microstructure of a granular material is quantified through a second-rank contact fabric tensor $\boldsymbol{A}$, which provides a convenient description of the directional distribution of the particle contact network and inherent structural anisotropy~\citep{oda1982,kenichi1984}. The orientational distribution $P(\boldsymbol{n})$ of contact normal unit vectors $\boldsymbol{n}$ can be expanded to the second order in Fourier series as~\citep{Rothenburg:1989iq,Bathurst:1990bw}:
\begin{equation}
    P(\boldsymbol{n})=\frac{1}{4\pi}\left[1+\boldsymbol{A}:\left(\boldsymbol{n}\otimes\boldsymbol{n}\right)\right],
    \label{eq21}
\end{equation}
where $\boldsymbol{A}$ is trace-free and symmetric. For dense granular materials where internal stress $\boldsymbol{\sigma}$ is dominated by particle contacts, it can be expressed as the following integral in the orientational space $\Omega$~\citep{Rothenburg:1989iq,Bathurst:1990bw,srivastava2019evolution}:
\begin{equation}
    \sigma_{ij}=\frac{N_{c}\langle l_{0} \rangle \langle f_{i} \rangle}{V}\int_{\Omega} P(\boldsymbol{n}) n_{j} \mathrm{d}\boldsymbol{n},
    \label{eq22}
\end{equation}
where $\langle l_{0} \rangle$ and $\langle f_{i} \rangle$ are the average magnitudes of the branch vector and the $i$-th component of the normal force between two contacting particles respectively. This representation of the stress tensor ensures that $\boldsymbol{\sigma}$ and $\boldsymbol{A}$ have the same structure. 

\subsection{Second Normal Stress Difference}
\begin{figure}
  \centerline{\includegraphics[width=0.95\columnwidth]{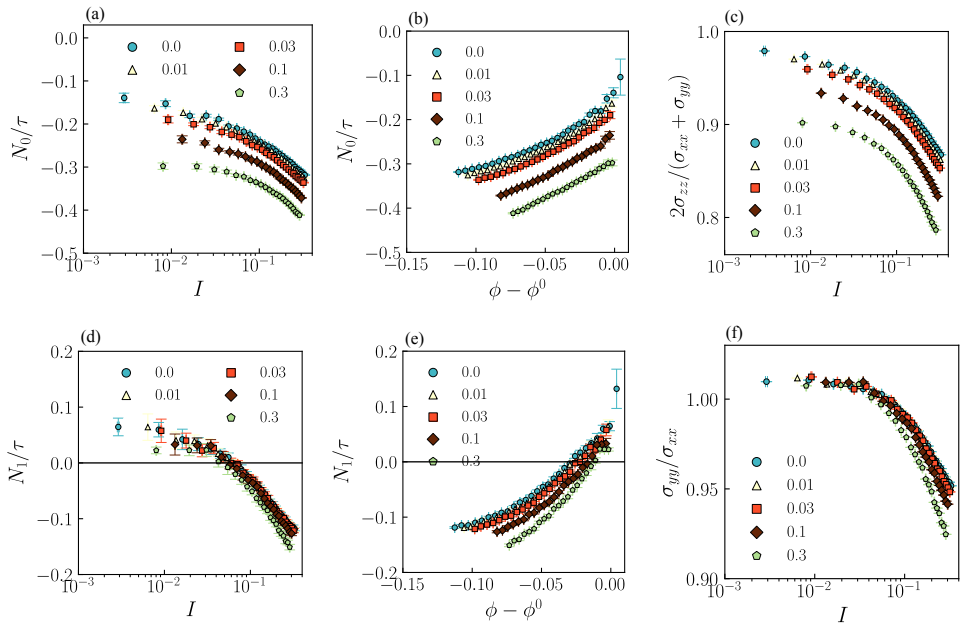}}
  \caption{Variation of scaled second normal stress difference $N_{0}/\tau$ with inertial number (a) $I$ and (b) distance to quasi-static solid volume fraction $\phi-\phi_{0}$, for five interparticle friction $\mu_{s}$ (see legend) at applied pressure $p_{\mathrm{ext}}=10^{-4}$. (c) Variation of $2\sigma_{zz}/\left(\sigma_{xx}+\sigma_{yy}\right)$ with $I$ for five interparticle friction. Variation of scaled first normal stress difference $N_{1}/\tau$ with inertial number (d) $I$ and (e) distance to quasi-static solid volume fraction $\phi-\phi_{0}$, for five interparticle friction $\mu_{s}$ (see legend) at applied pressure $p_{\mathrm{ext}}=10^{-4}$. (f) Variation of $\sigma_{yy}/\sigma_{xx}$ with $I$ for five interparticle friction.}
\label{fig6}
\end{figure}

Significant normal stress anisotropy emerges from the difference between the mean normal stress in the flow plane and normal stress in the vorticity direction, and is represented by the viscometric flow function $N_{0}/\tau=\left(2\sigma_{zz}-\sigma_{xx}-\sigma_{yy}\right)/2\tau$~\citep{seto2018}, where $x$ is the flow direction, $y$ is the flow gradient direction and $z$ is the vorticity direction, and the stresses are defined positive in the \textit{compressive} sense, since the forces are all repulsive. In the present simulations, $N_{0}/\tau$ is computed by:
\begin{equation}
\frac{N_{0}}{\tau} = \frac{-3\boldsymbol{\sigma}:\left(\boldsymbol{D}^{2}-\frac{\mathrm{tr}\left(\boldsymbol{D}^{2}\right)}{3}\boldsymbol{I}\right)}{\dot{\gamma}\boldsymbol{\sigma}:\boldsymbol{D}},
    \label{eq23}
\end{equation}
which is equivalent to $N_{0}/\tau=-\mu_{2}/\mu_{1}$. In figure~\ref{fig6}(\emph{b}), $N_{0}/\tau$ is plotted as a function of the distance to quasi-static solid volume fraction $\phi-\phi_{0}$ for five $\mu_{s}$. The negative value of $N_{0}$ implies that there is more normal stress in the flow plane than in the vorticity direction, as seen in figures~\ref{fig6}(\emph{a}-\emph{b}), and the ratio of the two normal stresses is consistent with previous simulations on dry granular flows (c.f. figure~\ref{fig6}(\emph{c}))~\citep{silbert2001,weinhart2013}. In present simulations, an imbalance between the external applied pressure and the internal pressure drives isochoric periodic cell deformation, as shown by the equal values of $F_{ii}(t)$ in figure~\ref{fig1}(\emph{b}). In another scenario where each $\sigma_{ii}$ is individually balanced, we observed a rapid compaction of the cell in the vorticity direction leading to simulation instability arising from the second normal stress difference. The magnitude of second normal stress difference is larger for frictional particles than for frictionless particles; however, even frictionless particles exhibit non-zero second normal stress difference during flow at finite inertial numbers, as seen in figure~\ref{fig6}(\emph{a}). As the solid volume fraction increases towards quasi-static $\phi_{0}$, the anisotropy consistently decreases for all $\mu_{s}$. For frictionless particles, $N_{0}$ appears to tend to zero in the quasi-static limit corresponding to $\phi^{0}\!=\!0.63$, which is similar to the random close packing density for mono-disperse spheres. However, the out of flow plane stress anisotropy is demonstrably non-zero for frictional particles even in the quasi-static limit, as also observed previously by \citet{seto2018}. The notion of non-zero anisotropy in the quasi-static regime is also consistent with recent observations of an anisotropic yield surface in frictional granular materials~\citep{thornton2010,li2012}.

An implication of these findings is that the flow of frictional granular materials is not co-directional, i.e., the hypothesis $\boldsymbol{\sigma}\!\propto\! \boldsymbol{D}$, which has been been assumed within the $\mu(I)$ rheological model is not accurate. $N_{0}/\tau$ increases with $I$ for all $\mu_{s}$, as seen in figure~\ref{fig6}(\emph{a}), and remains measurably non-zero for frictional particles even at low $I$. Prior simulations on quasi-static simple shear granular flows~\citep{sun2011}, granular flows down an incline~\citep{silbert2001,weinhart2013}, gravity-driven granular flows through an orifice~\citep{rycroft2009}, and granular flows in a split-bottom Couette cell~\citep{depken2007} have questioned the co-directionality hypothesis. Two previously proposed theoretical models---double shearing~\citep{spencer1964} and shear-free sheets~\citep{depken2007}---have also incorporated these effects for quasi-static and dense granular flows.

\begin{figure}
  \centerline{\includegraphics[width=0.95\columnwidth]{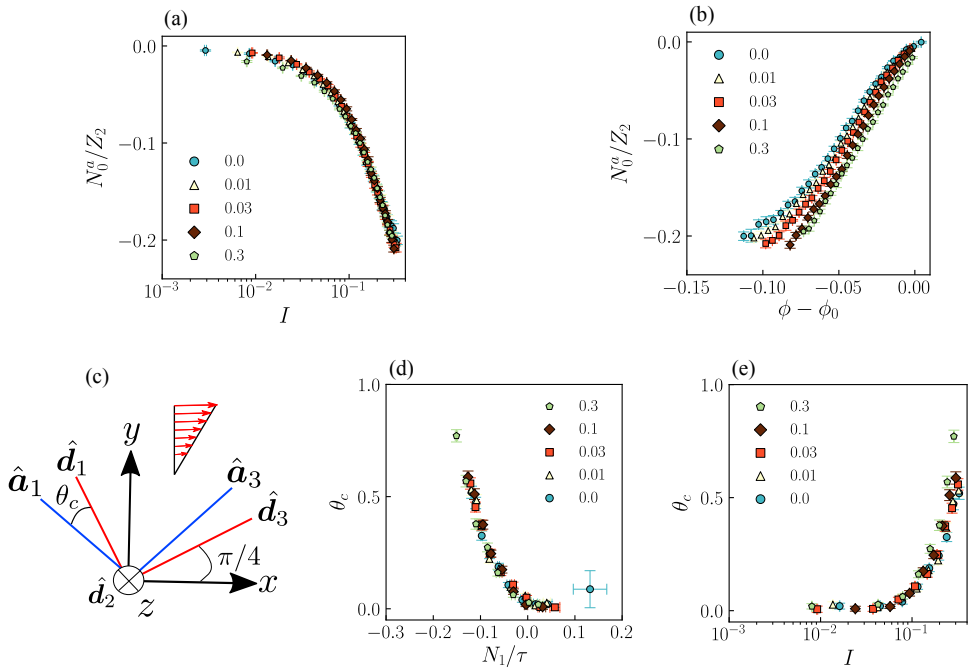}}
  \caption{Variation of contact fabric `second normal difference' $N_{0}^{a}$ scaled by rattler-free coordination $Z_{2}$ with (a) inertial number $I$ and (b) distance to quasi-static solid volume fraction $\phi-\phi_{0}$ for the five $\mu_{s}$ (see legend) at applied pressure $p_{\mathrm{ext}}=10^{-4}$. (c) A schematic depicting the misalignment angle $\theta_{c}$ between the principal directions of $\boldsymbol{D}$ and $\boldsymbol{A}$ in the flow plane (shown in red). Variation of $\theta_{c}$ with (d) $N_{1}/\tau$ and (e) inertial number $I$ for the five $\mu_{s}$ (see legend) at applied pressure $p_{\mathrm{ext}}=10^{-4}$. The vertical and horizontal error bars in (a)-(b) and (d)-(e) represent the standard error in the calculations.}
\label{fig7}
\end{figure}

The second normal stress difference results from an excess of contacts oriented in the flow plane than those oriented in the vorticity direction. Figure~\ref{fig7}(\emph{b}) shows the variation of $N_{0}^{a}/Z_{2}$ with $\phi-\phi_{0}$ for various interparticle friction. Here, $N_{0}^{a}=\left(A_{zz} - \frac{A_{xx}+A_{yy}}{2}\right)$ is the contact fabric `second normal difference', which represents the anisotropy in average orientation of contacts between the flow plane and the vorticity direction. The rattler-free coordination number is computed as $Z_{2}=2N_{c}/(N-N_{r})$, where $N_{r}$ is the number of rattler particles with less than two contacts, and $N_{c}$ is the total number of contacts with non-zero normal force belonging to non-rattler particles~\citep{sun2011}. At high $I$ corresponding to low $\phi$, a higher fraction of contacts are oriented in the flow plane, which results in large normal stress difference, as shown figure~\ref{fig7}(\emph{a}). Furthermore, all the data collapses onto a single curve for all inter-particle friction. Upon approach to the quasi-static regime at high $\phi$, the contact distribution becomes more isotropic, resulting in reduced normal stress difference. For frictionless particles, the orientational distribution of contacts expectedly becomes isotropic in the quasi-static regime at random close packing volume fraction, as seen by $N_{0}^{a} \to 0$. 

\subsection{First Normal Stress Difference}
The first normal stress difference, which characterizes the anisotropy between $\boldsymbol{\sigma}$ and $\boldsymbol{D}$ in the flow plane, is represented by the viscometric flow function $N_{1}/\tau\!=\!\left(\sigma_{yy}-\sigma_{xx}\right)/\tau$~\citep{guazzelli2018}. In the present simulations, this is estimated from:
\begin{equation}
\frac{N_{1}}{\tau} = \frac{\boldsymbol{\sigma}:\left(\boldsymbol{D}\boldsymbol{W}-\boldsymbol{W}\boldsymbol{D}\right)}{\dot{\gamma}\boldsymbol{\sigma}:\boldsymbol{D}},
    \label{eq24}
\end{equation}
which is equivalent to $N_{1}/\tau = 4\mu_{3}/\mu_{1}$. The variation of $N_{1}/\tau$ with $\phi-\phi_{0}$ for five $\mu_{s}$ is displayed in figure~\ref{fig6}(\emph{e}). At low $\phi$, $N_{1}$ is negative for all $\mu_{s}$, and its magnitude increases with increasing $\mu_{s}$ for a given distance from the quasi-static solid volume fraction $\phi-\phi^{0}$. At low $\phi$, the ratio of $N_{0}/N_{1}$ is approximately $3-4$ for all interparticle friction, consistent with previous findings~\citep{gallier2014}. The stress anisotropy in the flow plane increases with inertial number, as shown in figure~\ref{fig6}(\emph{d}), and also by the ratio of the two normal stresses in the flow plane, as shown in figure~\ref{fig6}(\emph{f}). The value of this ratio is consistent with previous simulations (c.f. figure~\ref{fig6}(\emph{f})) on granular flows down an incline~\citep{silbert2001,weinhart2013}.

When the flow becomes dense, $N_{1}$ increases towards zero and becomes slightly positive for highly dense flows in the quasi-static regime. The change of sign of $N_{1}$ at high $\phi$ has been previously observed in simulations~\citep{alam2005,weinhart2013,seto2018} and experiments~\citep{couturier2011}, but its existence is debated, and has been attributed to interparticle friction~\citep{dbouk2013} and boundary wall effects in experiments~\citep{gallier2014}. In the present simulations, we observe slightly positive $N_{1}$ for all values of $\mu_{s}$ at large $\phi$, and there are no boundary effects in these bulk simulations. Recently it was demonstrated that finite particle stiffness---which is often used as numerical regularization in hard particle simulations---causes $N_{1}$ to become positive at large $\phi$ in simulations on inertia-less frictional suspensions~\citep{seto2018}. However, our granular simulations do not provide any conclusive evidence of vanishing positive $N_{1}$ as a result of increasing particle stiffness. A careful analysis about this effect constitutes a part of our future work.

The first normal stress difference is related to the angular misalignment $\theta_{c}$ between the principal directions of $\boldsymbol{D}$ ($\hat{\boldsymbol{d}_{1}}$ and $\hat{\boldsymbol{d}_{3}}$) and $\boldsymbol{A}$ ($\hat{\boldsymbol{a}}_{1}$ and $\hat{\boldsymbol{a}}_{3}$) in the flow plane, as described in the schematic in figure~\ref{fig7}(\emph{c}). Here, $\hat{\boldsymbol{d}}_{1}$ and $\hat{\boldsymbol{d}}_{3}$ represent the compression and expansion directions of shear flow respectively, and $\theta_{c}$ represents the angle between $\hat{\boldsymbol{d}}_{1}$ and the major principal direction $\hat{\boldsymbol{a}}_{1}$ of $\boldsymbol{A}$. The misalignment angle $\theta_{c}$ and $N_{1}/\tau$ are strongly correlated, as depicted in figure~\ref{fig7}(\emph{d}) for various $\mu_{s}$ that largely collapse onto a single curve, indicating a one-to-one correspondence between stress anisotropy and the misalignment~\citep{seto2018}. The misalignment between $\boldsymbol{A}$ and $\boldsymbol{D}$ results in excess stress along the flow direction as compared to the gradient direction, which sets the negative sign of first normal stress difference, similar to previous observations in dry granular flows~\citep{silbert2001,weinhart2013}. Such microstructural origins of $N_{1}$ arising from a misalignment between the projected contact vectors and principal flow direction in the flow plane were previously demonstrated for inertia-less frictional suspensions~\citep{seto2018}, and in the case of dilute granular flows through a similar misalignment between fluctuating velocity moment tensor and principal flow direction in the flow plane~\citep{saha2016}. Remarkably, $\theta_{c}\to 0$ as $N_{1}\to 0$, indicating a vanishing misalignment between $\boldsymbol{D}$ and $\boldsymbol{A}$ at high solid volume fractions. This is also seen by the variation of $\theta_{c}$ with $I$ in figure~\ref{fig7}(\emph{e}), where the data for all $\mu_{s}$ collapse onto a single curve. A small positive first normal stress difference exists at high solid volume fractions in the vicinity of yield stress despite a near-complete alignment of $\boldsymbol{D}$ and $\boldsymbol{A}$ in the flow plane, thus indicating that either a different underlying physical phenomenon is responsible for positive $N_{1}$, or it is possibly a consequence of finite system size.

The observations of microstructure-induced normal stress differences in dense granular flows, especially the collapse of $N_{0}^{a}/Z_{2}$ and $\theta_{c}$ with $I$ in figures~\ref{fig7}(\emph{a} and \emph{e}), indicate that fabric tensor is an appropriate internal state variable that can be used to construct a rheological model with evolution equations for the microstructure, as was done for quasi-static granular flows~\citep{sun2011,parra2019}. Such an approach has been previously used in modeling suspension rheology~\citep{goddard2006,stickel2006}, and a general framework adaptable to dry granular flows has been provided by ~\citet{goddard2014}.

\section{Conclusions}
In this paper, we described a discrete element method to simulate dense granular flows under external applied stress in a fully periodic representative volume element. Rather than prescribing solid volume fraction and/or strain rate, this method enables independent evolution of solid volume fraction and 3D strain rate tensor in response to an imbalance between internal state of stress and external applied stress. Using this method, bulk viscometric granular flows were simulated under external pressure and shear stress, which was devoid of any boundary effects, and thus closely represented the boundary conditions often found in practice.

We developed a second-order rheological model to relate the internal Cauchy stress $\boldsymbol{\sigma}$ with the strain rate tensor $\boldsymbol{D}$ for various interparticle friction. The model considers both rate-dependent and rate-independent contributions to the total stress, where the latter is often described using models of granular plasticity. The rheological model well-predicts the $\mu(I)$ rheology of granular materials. Additionally, it also predicts normal stress differences in steady viscometric granular flows, which have often been observed in simulations and experiments, but have not been well-characterized. A major implication of this model is that it does not impose co-axiality between $\boldsymbol{\sigma}$ and $\boldsymbol{D}$ in dense granular flows, which is often assumed in several other constitutive models.

A major focus of this work has been to highlight the role of interparticle friction on viscometric granular rheology in the dense flowing regime, particularly on the two normal stress differences. We found that friction not only increases the quasi-static shear stress ratio, but also the quasi-static value of the second normal stress difference, thus indicating the presence of an anisotropic yield stress, whereas frictionless particle do not exhibit such anisotropy in the quasi-static regime at solid volume fraction similar to the random close packing of monodisperse spheres. At higher flow rates in the inertial regime, friction consistently increases the magnitude of both normal stress differences, indicating an increasing departure from the co-axiality of $\boldsymbol{\sigma}$ and $\boldsymbol{D}$. Although the second normal stress difference is always negative, the first normal stress difference changes sign from negative to positive at high solid volume fractions in the quasi-static regime. Further microstructural investigations highlighted that negative first normal stress difference results from a misalignment between $\boldsymbol{D}$ and a second-rank contact fabric tensor $\boldsymbol{A}$ in the flow plane, which describes the orientational distribution of sphere-sphere contacts in granular flows. Furthermore, the magnitude of misalignment increases with the inertial number similarly for all inter-particle frictions. The second normal stress difference results from an excess of contacts oriented in the flow plane than in the vorticity direction, which is also observed from the anisotropy in the normal components of the fabric tensor. Upon appropriate normalization with friction-dependent coordination number, the fabric tensor anisotropy was shown to collapses onto a single curve for all inter-particle frictions.

These results demonstrate the importance of developing rheological models beyond simple scalar models to predict granular rheology in even simple shear flows, and certainly for complex and heterogeneous flow fields that are observed in practice. The breakdown of co-axiality of stress and strain rate tensors highlights the need for an anisotropic rheological model that includes contact fabric tensor as an internal variable. A general form of such anisotropic models $\boldsymbol{\sigma}=\mathcal{F}(\boldsymbol{D},\boldsymbol{A})$ was recently proposed for granular materials and suspensions~\citep{goddard2006,stickel2006,goddard2014}, and was calibrated for rate-independent granular flows~\citep{sun2011,parra2019}. The calibration of these models in rate-dependent flows is required, along with other non-viscometric flows such as uniaxial or triaxial compression, as well as transient evolving inertial granular flows. These topics are currently a subject of our ongoing study. Lastly, the constitutive model described here could also be extended to include other important granular flow phenomena such as hysteresis~\citep{degiuli2017,perrin2019} and non-locality~\citep{henann2013} at low inertial numbers.

\section{Acknowledgements}\label{ack}
The authors acknowledge helpful discussions with D. Henann and K. Kamrin. This work was performed at the Center for Integrated Nanotechnologies, a U.S. Department of Energy and Office of Basic Energy Sciences user facility. Sandia National Laboratories is a multimission laboratory managed and operated by National Technology and Engineering Solutions of Sandia, LLC, a wholly owned subsidiary of Honeywell International, Inc., for the U.S. Department of Energy's National Nuclear Security Administration under Contract No. DE-NA-0003525. This paper describes objective technical results and analysis. Any subjective views or opinions that might be expressed in the paper do not necessarily represent the views of the U.S. Department of Energy or the United States Government.

\noindent Declaration of Interest: The authors report no conflict of interest.

 \appendix

 \section{Fitting Parameters of the Rheological Model}

 In this Appendix, Table 1 provides the fitting parameters $A_{1}$, $A_{2}$, $A_{3}$ and $A_{4}$ of the rheological model, defined in equations (\ref{eq15}), (\ref{eq17}), (\ref{eq19}) and (\ref{eq20}) respectively, as a function of interparticle friction $\mu_{s}$. 

\begin{table}
    \centering
    \label{tab1}
    \begin{tabular}{lllll}
        \toprule
        $\mu_{s}$ & $A_1$ & $A_2$ & $A_3$ & $A_4$  \\
        \midrule
0.0 & 0.530 & 0.240 & 0.098 & 0.272\\
0.001 & 0.530 & 0.243 & 0.098 & 0.274\\
0.003 & 0.528 & 0.247 & 0.102 & 0.276\\
0.01 & 0.522 & 0.252 & 0.107 & 0.274\\
0.03 & 0.516 & 0.270 & 0.113 & 0.263\\
0.04 & 0.516 & 0.274 & 0.116 & 0.259\\
0.05 & 0.520 & 0.284 & 0.118 & 0.254\\
0.06 & 0.518 & 0.290 & 0.128 & 0.251\\
0.08 & 0.531 & 0.322 & 0.127 & 0.247\\
0.1 & 0.526 & 0.303 & 0.141 & 0.246\\
0.14 & 0.545 & 0.330 & 0.140 & 0.243\\
0.18 & 0.554 & 0.360 & 0.153 & 0.241\\
0.23 & 0.562 & 0.366 & 0.151 & 0.238\\
0.3 & 0.573 & 0.398 & 0.151 & 0.230\\
        \bottomrule
    \end{tabular}
    \caption{Fitting parameters corresponding to equations (\ref{eq15}), (\ref{eq17}), (\ref{eq19}) and (\ref{eq20}), as a function of interparticle friction $\mu_{s}$.}
\end{table}

\bibliographystyle{jfm}

\end{document}